\documentclass[12pt,letterpaper]{JHEP3}

\usepackage{bm}
\usepackage{graphicx}
\usepackage{amsmath}
\usepackage{amssymb}

\def\bR{\mathbb{R}}
\def\bZ{\mathbb{Z}}
\def\cN{\mathcal{N}}
\def\SO{\mathrm{SO}}
\def\USp{\mathrm{USp}}
\def\SU{\mathrm{SU}}
\def\U{\mathrm{U}}
\def\CP{\mathbb{CP}}
\def\inc#1#2{{%
\setbox1=\hbox{\includegraphics[scale=#1]{#2}}%
\raise-.5\ht1\box1%
}}
\def\sizeA{.28}
\def\sizeB{.35}
\def\sizeC{.5}

\title{\boldmath
Six-dimensional $D_N$ theory and\\
four-dimensional $\SO$--$\USp$ quivers}
\author{
Yuji Tachikawa \\

School of Natural Sciences, Institute for Advanced Study,\\
 Princeton,  New Jersey 08540, USA 
}

\preprint{}
\abstract{
We realize four-dimensional $\cN=2$ superconformal quiver gauge theories 
with alternating $\SO$ and $\USp$ gauge groups 
as compactifications of the six-dimensional $D_N$ theory with defects.
The construction can be used to analyze infinitely strongly-coupled limits 
and S-dualities of such gauge theories,
resulting in a new class of isolated four-dimensional $\cN=2$ superconformal field theories
with $\SO(2N)^3$ flavor symmetry.
}

\begin{document}

\section{Introduction}

A method to systematically understand and construct a large class 
of four-dimensional $\cN=2$ superconformal field theories (SCFTs) 
was recently presented by Gaiotto
in \cite{Gaiotto:2009we}.
By means of a clever  rewriting of the known Seiberg-Witten curves for
quiver  theories based on  $\SU$ gauge groups \cite{Witten:1997sc},
Gaiotto showed that such a theory 
arises as a compactification on a Riemann surface
of the six-dimensional $A_{N-1}$ theory with $(2,0)$ supersymmetry, 
with punctures associated to defect operators.
The marginal couplings of a quiver theory are encoded in the moduli of the
punctured Riemann surface, and both weakly-coupled and
strongly-coupled limits were shown to correspond to degenerations
of the Riemann surface.
This  approach gave a unified perspective on the S-dualities of 
$\SU(2)$ gauge theory with four flavors \cite{Seiberg:1994aj},
which involved the triality of $\SO(8)$ flavor symmetry,
and of $\SU(3)$ gauge theory with six flavors,
the strongly-coupled limit of which is dual to the mysterious isolated SCFT with $E_6$ flavor symmetry \cite{Minahan:1996fg} coupled to
an $\SU(2)$ gauge group with one flavor  \cite{Argyres:2007cn}.
It also predicted a whole new family of  SCFTs with $\SU(N)^3$ flavor symmetry
which are isolated, i.e.~have no marginal couplings.
The holographic description of these theories
was found and discussed  in \cite{Gaiotto:2009gz}.

We call this Riemann surface with punctures 
the {\em G-curve} of the theory, in order to avoid confusion and
to distinguish it from the Seiberg-Witten curve, which is, roughly speaking, 
an $N$-sheeted cover of the G-curve. 
In this framework, vacuum expectation values (vevs) of 
dimension-$d$ Coulomb branch operators
are encoded in a degree-$d$ differential on the Riemann surface, 
which is allowed to have poles at each of the punctures.
These differentials are the scalar fields of the six-dimensional $A_{N-1}$ theory.
Therefore, from the six-dimensional viewpoint,
each puncture corresponds to a defect operator that introduces singularities to fields, 
much like 't Hooft loops or surface operators do in four-dimensional gauge theory.

It was observed 
in \cite{Gaiotto:2009we} that the punctures of the $A_{N-1}$ theory
are naturally labeled
by Young tableaux with $N$ boxes, which also specify
embeddings of $\SU(2)$ into $\SU(N)$.
It was also found that there is a natural mapping between
tableaux and tails of conformal quivers.
In other words, we can obtain information about the elusive six-dimensional
conformal field theory from the study of the quiver theories with $\SU$ gauge groups.

The main objective of this paper is to repeat Gaiotto's analysis 
for the quivers with $\USp$ and $\SO$ gauge groups, by realizing them
using M5-branes at an M-theory orientifold.
Recall that six-dimensional $A_{N-1}$ theory is realized
as the low-energy effective theory on the $N$ coincident M5-branes;
one can then introduce M-theoretic orientifolding, 
which flips five directions of the spacetime.
$2N$ M5-branes on top of the orientifold singularity realize in the low energy limit
the six-dimensional $D_N$ theory.

We first show that the Seiberg-Witten curves of these quivers
can be recast into a form which makes manifest their correspondence
to compactifications of the $D_N$ theory on Riemann surfaces with punctures.
We construct new isolated SCFTs with $\SO(2N)^3$ flavor symmetry,
which appears when the Riemann surface on which 
the $D_N$ theory is compactified degenerates and develops a sphere 
with three  necks attached.

We then study defects of the $D_N$ theory via an analysis
of the tails of superconformal quivers. We find that defects are naturally
labeled by embeddings of $\SU(2)$ into either $\SO(2N)$ or $\USp(2N-2)$.

We will see, along the way, that the compactification of the $A_3$ theory
and of the $D_3$ theory on the same surface 
gives the same four-dimensional theory.  
The way it works is rather nontrivial: in the four-dimensional
description the $A_3$ theory involves hypermultiplets in the 
$\mathbf{4}$ of $\SU(4)$ while 
the $D_3$ theory contains multiplets in the $\mathbf{6}$ of $\SO(6)$.
In the M-theory description, the $A_3$ theory is defined on
a stack of four M5-branes, whereas the $D_3$ theory is realized by
six M5-branes on the M-theory orientifold. We will find that 
subtle properties of the M-theory orientifold \cite{deBoer:1998by,Hori:1998iv}
play crucial roles in this equivalence. 
All these facts support non-perturbative equivalence of
the $A_3$ theory and the $D_3$ theory
as six-dimensional superconformal theories.

Finally we will see that the isolated SCFT with $E_7$ flavor symmetry\cite{Minahan:1996cj}
arises from a strongly-coupled limit of a particular quiver with a $\USp(4)$ factor,
as anticipated in \cite{Argyres:2007cn}.

The paper is organized as follows: we  start by reviewing the analysis of 
$\SU(N)$ quivers and their relation to the $A_{N-1}$ theory in Sec.~\ref{Atype}.
We then analyze the $\SO$--$\USp$ quivers and their relation to the $D_N$ theory
in Sec.~\ref{Dtype}. 
We conclude with some  discussion in Sec.~\ref{discussion}.
There are two appendices:
Appendix~\ref{so4} is an analysis of $\SO(4)$--$\USp(2)$ quivers in our framework.
Appendix~\ref{curves} contains a detailed derivation 
of the G-curve of $\SO$--$\USp$ quivers
from the corresponding Seiberg-Witten curve.

\section{Review: 6d $A_{N-1}$ theory and $\SU(N)$ quivers} \label{Atype}
\subsection{Superconformal quivers, G-curve and Young tableaux}
Let us start by considering  an $\cN=2$ supersymmetric linear quiver gauge theory
with a chain of  $\SU$ groups
\begin{equation}
\SU(d_1)\times \SU(d_2) \times\cdots \times \SU(d_{n-1}) \times \SU(d_n),
\end{equation}  a bifundamental hypermultiplet between each pair of consecutive gauge groups 
$\SU(d_a)\times \SU(d_{a+1})$,
and $k_a$ extra fundamental hypermultiplets for $\SU(d_a)$.
To make every gauge coupling constant marginal, we require
\begin{equation}
k_a= 2da-d_{a-1}-d_{a+1} = (d_{a}-d_{a-1}) -(d_{a+1}-d_{a}),\label{SUmarginalitycondition}
\end{equation}   where we defined $d_0=d_{n-1}=0$.
Since $k_a$ is non-negative, we have
\begin{equation}
d_1 < d_2 <  \cdots  < d_l  = \cdots = d_r > d_{r+1} > \cdots > d_n. \label{generallinear}
\end{equation} 

Let us denote $N=d_l=\cdots =d_r$; 
we refer to the parts to the right of $d_r$ and to the left of $d_l$
as the two tails of this superconformal quiver.
Consider the tail on the right hand side of \eqref{generallinear}, \begin{equation}
N=d_r>d_{r+1}> \cdots > d_n. \label{SUtail}
\end{equation}
$d_a-d_{a+1}$ is monotonically non-decreasing because $k_a\ge 0$;
therefore we can associate naturally a Young tableau to the tail
by requiring that it has a row of width $d_a-d_{a+1}$ for each $a\ge r$.
For illustration, the possible types of tails for $N=4$ 
are shown in Table~\ref{SU4quivers}. As is customary, a circle or a box with $n$ inside
stands for an $\SU(n)$ gauge group or flavor symmetry respectively,
and a line connecting two objects stands for a bifundamental hypermultiplet.
\begin{table} 
\[
\begin{array}{@{\extracolsep{1ex}}c@{\hskip1ex}|c@{\hskip1ex}|c@{\hskip1ex}|c@{\hskip1ex}||c@{\hskip1ex}|c}
\text{Tableau} &\text{Alias}&  \text{Flavor} & \text{Poles} & \text{Quiver} & \text{G-curve}  \\
&&&(p_2,p_3,p_4) &&\\
\hline
\inc{\sizeB}{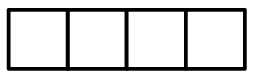} &\odot&  \SU(4) &  (1,2,3) & \inc{\sizeA}{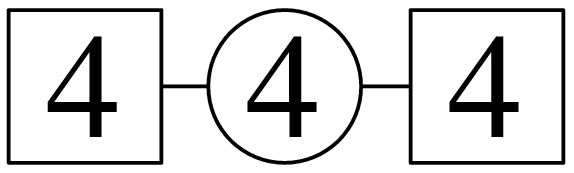} & \hbox{\vbox to 2em{}}\inc{\sizeA}{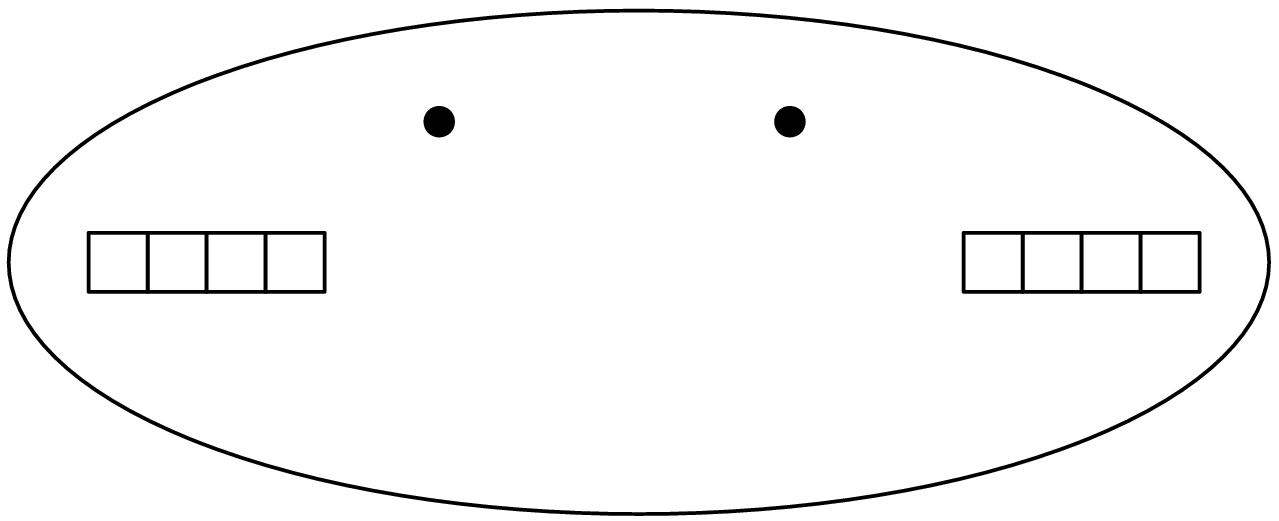}\vphantom{\Big|} \\[2em]
\inc{\sizeB}{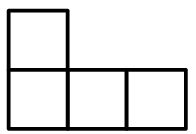} &&  \SU(2)\times \U(1)  & (1,2,2) & \inc{\sizeA}{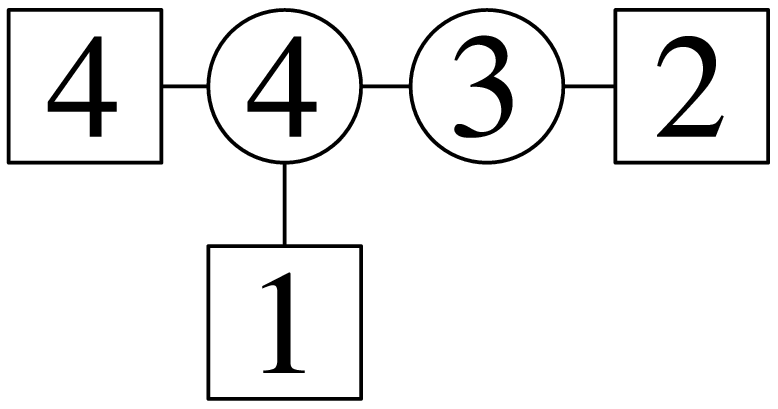} & \inc{\sizeA}{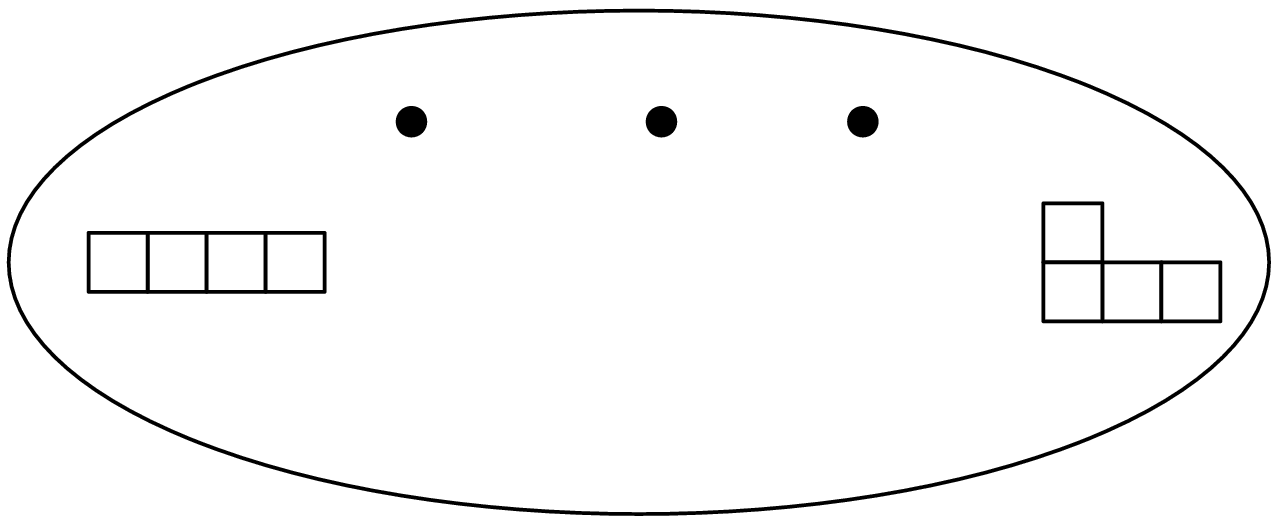} \\[2em]
\inc{\sizeB}{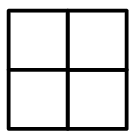} &&  \SU(2) & (1,1,2) &\inc{\sizeA}{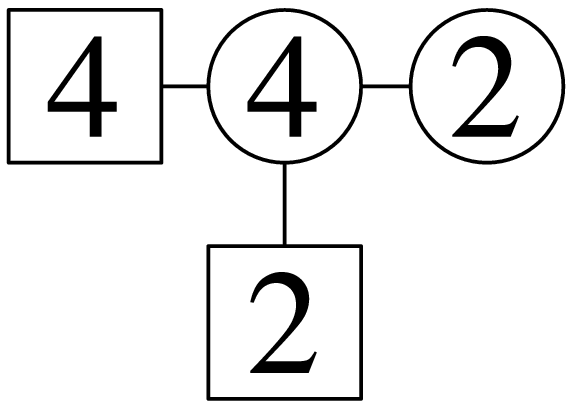} & \inc{\sizeA}{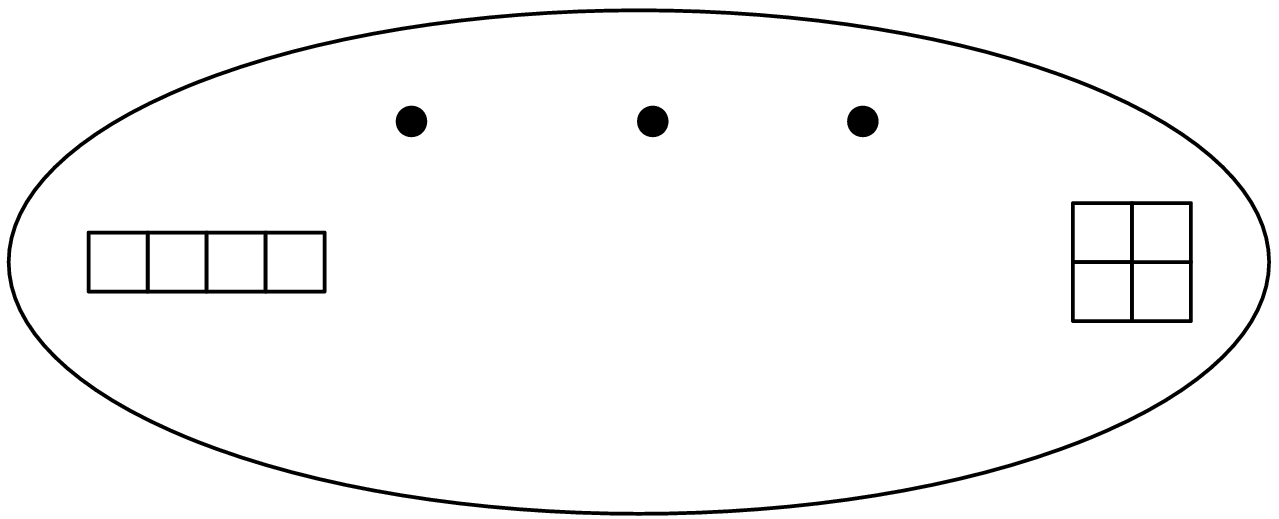} \\[2em]
\inc{\sizeB}{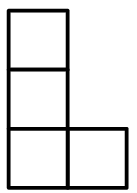} &\bullet& \U(1) &  (1,1,1) & \inc{\sizeA}{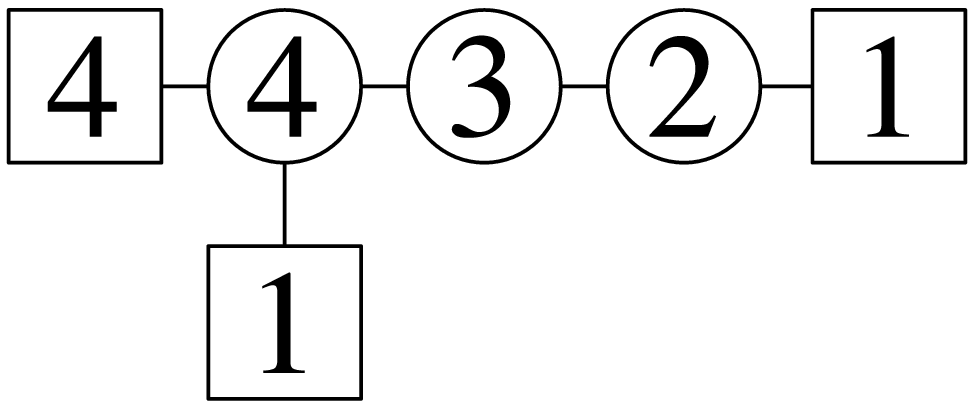} & \inc{\sizeA}{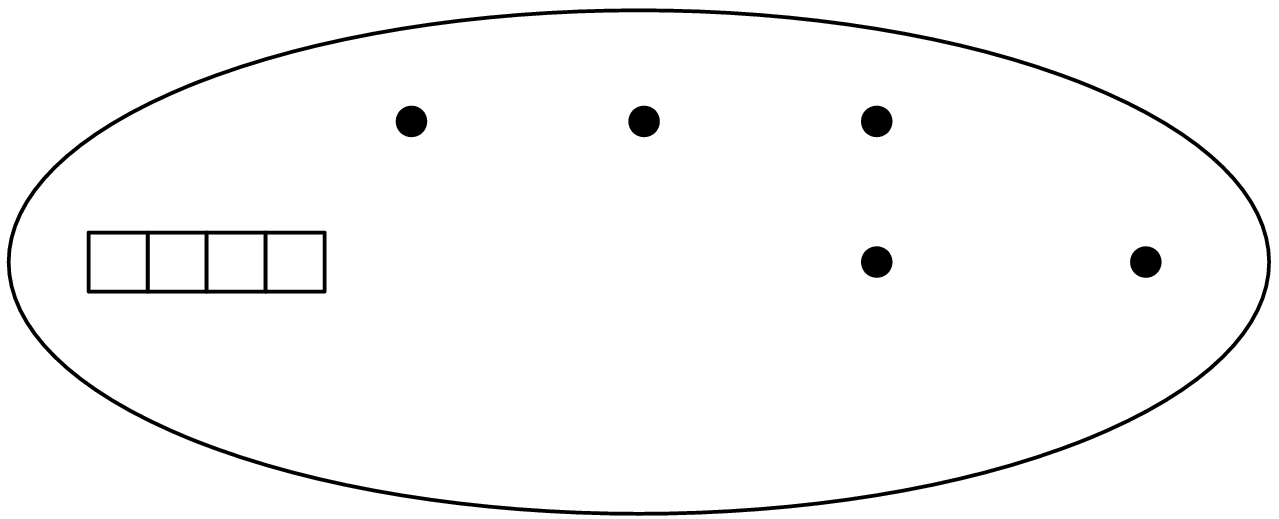} \\[2em]
\inc{\sizeB}{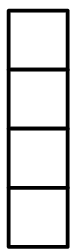} &&  \text{none}& (0,0,0) & \inc{\sizeA}{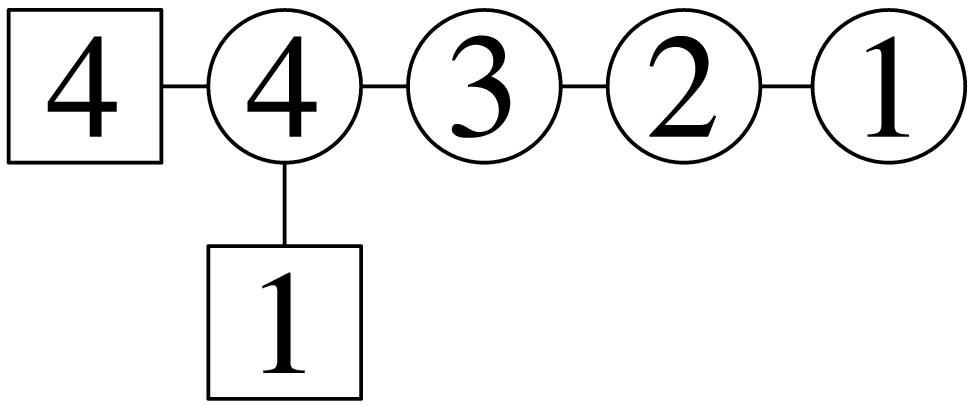} & \inc{\sizeA}{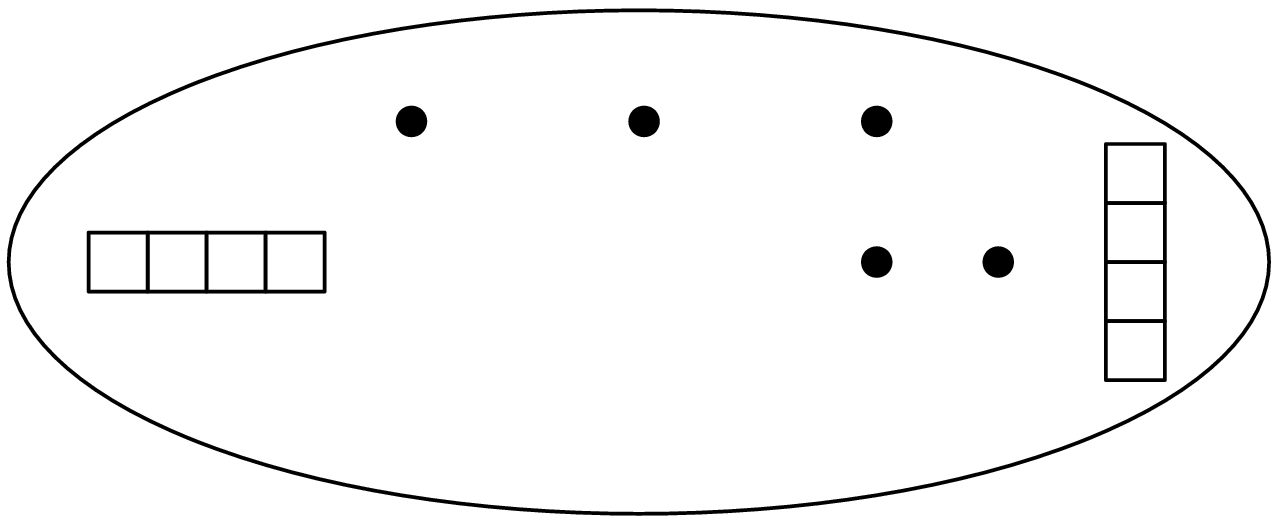}
\end{array}
\]
\caption{Young tableaux with $N$ boxes label the punctures of the six-dimensional $A_{N-1}$ theory. The case $N=4$ is shown here.
Each tableau has its associated flavor symmetry, 
and worldvolume fields $\phi_k$ are allowed to have poles of degree $p_k$
at the puncture.
A puncture whose tableau consists of one row of width $N$ is also known as
a {\it full puncture} and marked by $\odot$.
A puncture whose tableau consists of one column of height $N-1$ and 
another of height $1$ is  the same as 
a {\it simple puncture} marked by $\bullet$.
For each tableau, a four-dimensional quiver gauge theory with the corresponding tail is shown. 
 `$\SU(1)$' gauge groups need to be understood as a shorthand for
 the brane construction, as explained in the text. \label{SU4quivers}}
\end{table}

The Seiberg-Witten curves for these quivers were originally found in \cite{Witten:1997sc},
and rewritten into the following form in \cite{Gaiotto:2009we}:
We start from a Riemann surface $\Sigma$, in this case $\Sigma=\CP^1$,
with several punctures on it. 
Then the Seiberg-Witten curve is realized as a subspace
of the total space $T^*\Sigma$ of the bundle of holomorphic differential 
on $\Sigma$, given as follows:
 \begin{equation}
0=x^N+x^{N-2}\phi_2 + x^{N-3}\phi_3+ \cdots+ \phi_N\label{SWcurve}
\end{equation} where $x$ is a holomorphic differential on the Riemann surface $\Sigma$,
and $\phi_d$ is a degree-$d$ differential with poles at the punctures,
encoding vevs of Coulomb branch operators of dimension $d$.
Then the Seiberg-Witten differential is $x$ itself.
This form makes the superconformal property of the theory manifest:
one can assign the scaling dimension of the various fields to be equal to  the degree 
of the corresponding differentials.
One finds that, for the general quiver \eqref{generallinear},
one has $n+1$ punctures of the same type, which we call the {\em simple punctures} and denote by $\bullet$,
and two extra punctures each of which encodes the  type of the tails.
We label these two punctures by the Young tableaux associated to the corresponding tails.
In Table~\ref{SU4quivers}, the curves with punctures are shown along with the corresponding
quiver gauge theories.
The puncture whose tableau consists of one row of width $N$ is called  
the {\em full puncture} and labeled  by $\odot$.

As explained in \cite{Gaiotto:2009we},
this system can be thought of as a compactification of the  six-dimensional
$A_{N-1}$ $(2,0)$-theory on $\Sigma$ with defects at the punctures.
Recall that the $A_{N-1}$ theory is the low-energy limit of the theory of 
$N$ coincident M5-branes, or of the compactification of type IIB strings 
on the four-dimensional asymptotically locally Euclidean (ALE) space of type $A_{N-1}$;
this theory has operators of dimension $2,3,\ldots,{N}$.
Compactifications on a Riemann surface which preserves  the supersymmetry 
involve twisting as usual, which turns an operator of dimension $d$ into 
a meromorphic differential of degree $d$.
Another way to understand this twisting is to recall that the space in which the M5-branes
are embedded needs to be hyperk\"ahler to preserve $\cN=2$ supersymmetry
in four dimensions. The neighborhood of  $\Sigma$ in such a space 
can be approximated by $T^*\Sigma$, which is exactly the space used in \eqref{SWcurve}.
Then $N$ solutions of \eqref{SWcurve} determine the position of $N$ M5-branes
in the fiber direction $x$, at each point of the base $\Sigma$.

This construction generalizes the realization of  Seiberg-Witten curves 
as compactifications of the $A_{N-1}$ theory discussed
in \cite{Klemm:1996bj,Katz:1997eq}: the
$\phi_d$ are now allowed to have poles at a finite number of punctures on $\Sigma$.
These describe conformal defects of the $A_{N-1}$ theory.

At a simple puncture $\phi_d$ is allowed to have a simple pole.
The orders of poles $\phi_d$  at a general puncture 
can be determined from the Seiberg-Witten curve
of the corresponding superconformal quiver, and 
can be easily read off from the associated tableau.
Given a tableau with rows of width $w_1 \ge w_2 \ge w_3 \cdots$,
we define a sequence of integers $\nu_i$ as follows \begin{equation}
(\nu_1,\nu_2,\ldots)= (\underbrace{1,\ldots,1}_{w_1},\underbrace{2,\ldots,2}_{w_2},\cdots,).
\end{equation}
Then, $\phi_d$ is allowed to have poles of order $p_d=d-\nu_d$.
The set of orders $p_d$ define the {\it pole structure} of the  puncture.
Again for illustration, the tableaux and the corresponding pole structures $(p_2,p_3,p_4)$ 
are listed in Table~\ref{SU4quivers} for the case $N=4$.

One notable property is that the puncture associated to the tail whose
tableau consists of  one column of height $N-1$ and another of height $1$ 
has the same pole structure as the simple puncture.  
Furthermore, the rightmost gauge group of the tail
is $\SU(2)$ coupled effectively to four flavors,
and the S-duality of this gauge group exchanges 
the puncture associated to the tail with the simple puncture.
We therefore identify the simple puncture with the puncture associated to this tableau.

We call this set consisting of a Riemann surface $\Sigma$ and punctures marked by 
Young tableaux the {\em G-curve} of the system, to distinguish it from the Seiberg-Witten curve.
Given pole structures at the punctures, the number of moduli in $\phi_d$
is the dimension of the space of  
the holomorphic differentials  of degree $d$ with prescribed singularities,
given by the formula
\begin{equation}
\text{\# moduli in $\phi_{d}$}
= (\sum_\text{punctures} p_d ^{(i)}) - (2d-1)\label{counting}
\end{equation}  
where $(p_d^{(i)})$ is the pole structure of the $i$-th puncture.
The use of this formula will be illustrated in Sec.~\ref{SUpants}.

In Table~\ref{SU4quivers}, the tableau with one column of height $4$ is also listed.
In general, a tableau with one column of height $N$
 does not apparently have a corresponding superconformal tail,
because the rule explained above would associate a tail of the form \begin{equation}
\cdots\times\SU(N)\times \SU(N-1)\times \cdots \times \SU(2) \times \text{``$\SU(1)$''}.
\label{su1tail}
\end{equation}
One also finds that none of the $\phi_d$ are allowed to have poles at the  `puncture'
corresponding to this tableau.

\begin{figure}
\[
\inc{\sizeB}{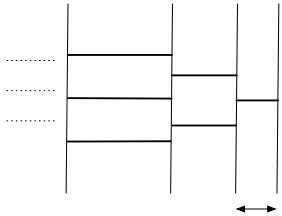}
\]
\caption{Brane configuration involving ``$\SU(1)$'' part. Vertical lines stand for
NS5 branes extended along $x^{0,1,2,3,4,5}$, and horizontal lines are D4-branes
extended along $x^{0,1,2,3,6}$ where the extent $x^6$ is bounded by 
two NS5-branes. It would correspond to a quiver with gauge groups
$\cdots\times\SU(3)\times\SU(2)\times \text{``$\SU(1)$''}$.\label{su1}}
\end{figure}

This sounds problematic, but by using a string-theoretic construction
one can make sense of it. Consider a brane configuration in type IIA string theory
shown in Fig~\ref{su1}. There, vertical lines represent NS5-branes
and horizontal lines D4-branes suspended between them, as discussed in \cite{Witten:1997sc}.
Thus this configuration shows the tail of a quiver with the gauge groups \eqref{su1tail}
where the ``$\SU(1)$ part'' just decouples in the infrared limit.
Still, this configuration can be lifted to a configuration of a connected M5-brane in M-theory.
Its rewriting  produces a Riemann surface with simple punctures
and one extra `puncture', at which none of $\phi_d$ has poles. 
This extra puncture is not totally devoid of physical meaning, as its position encodes the
separation of the last two NS5-branes, which  roughly corresponds to the `gauge coupling'
of the `$\SU(1)$ gauge group'.
Therefore we find that it is natural to associate the Young tableau with one column of height $N$ 
to this type of puncture.

\subsection{Punctures and associated flavor symmetries}
Let us now consider the flavor symmetry associated to a puncture labeled
by a given Young tableau.
As we saw, a tail \eqref{SUtail} of the $\SU(N)$ quiver gives
a number of simple punctures and a puncture associated to the Young tableau
with rows of width $d_r-d_{r+1}$, $d_{r+1}-d_{r+2}$, \ldots, $d_{n-1}-d_n$.
One finds that the $U(1)$ symmetry acting on each of the bifundamental hypermultiplets
is carried by the simple punctures,
and the flavor symmetries of $k_a$ fundamental hypermultiplets
of $\SU(d_a)$ gauge groups are associated to the punctures labeled by the tableau.
We can easily read off the flavor symmetry from a given tableau:
Let $l_h$  be the number of columns of height $h$.
Then, for each $l_h\ne 0$
there are $l_h$ fundamental hypermultiplets
coupled to one of the gauge groups in the tail, which gives $\U(l_h)$ symmetry.
The overall $\U(1)$ is carried by the simple puncture closest to the puncture
labeled by the tableau.
Therefore the flavor symmetry is given by \begin{equation}
\mathrm{S}\left[\prod_{l_h>0} \U(l_h)\right].\label{flavorSU}
\end{equation}
For the tails of $\SU(4)$ quivers, these flavor symmetries
are listed in Table~\ref{SU4quivers}.

Let us note one curious mathematical fact:
for a given tableau, we may associate  
an  embedding of $\SU(2)$ into $\SU(N)$ described by the decomposition
of the fundamental representation of $\SU(N)$ into the irreducible representations
of $\SU(2)$, given by 
\begin{equation}
N\to \underbrace{1+1+\cdots+1}_{l_1}+ 
\underbrace{2+\cdots+2}_{l_2}+ \cdots.
\end{equation}
Then, its commutant inside $\SU(N)$ is easily seen to agree with \eqref{flavorSU}.

\subsection{SCFT with $\SU(N)^3$ flavor symmetry}\label{SUpants}
With the interpretation of the G-curve as the compactification of the $A_{N-1}$ theory,
one can easily derive various new types of S-duality, generalizing the ones found in 
\cite{Argyres:2007cn}.
As an example, 
let us recall the construction of a class of  SCFTs 
with $\SU(N)^3$  flavor symmetry in \cite{Gaiotto:2009we},
whose gravity dual was found in \cite{Gaiotto:2009gz}.
The general method was
detailed in these papers, so we use a specific example of an $\SU(4)$ quiver
to illustrate the procedure.

\begin{figure}
\[
\begin{array}{c@{\qquad}c}
 \inc{\sizeA}{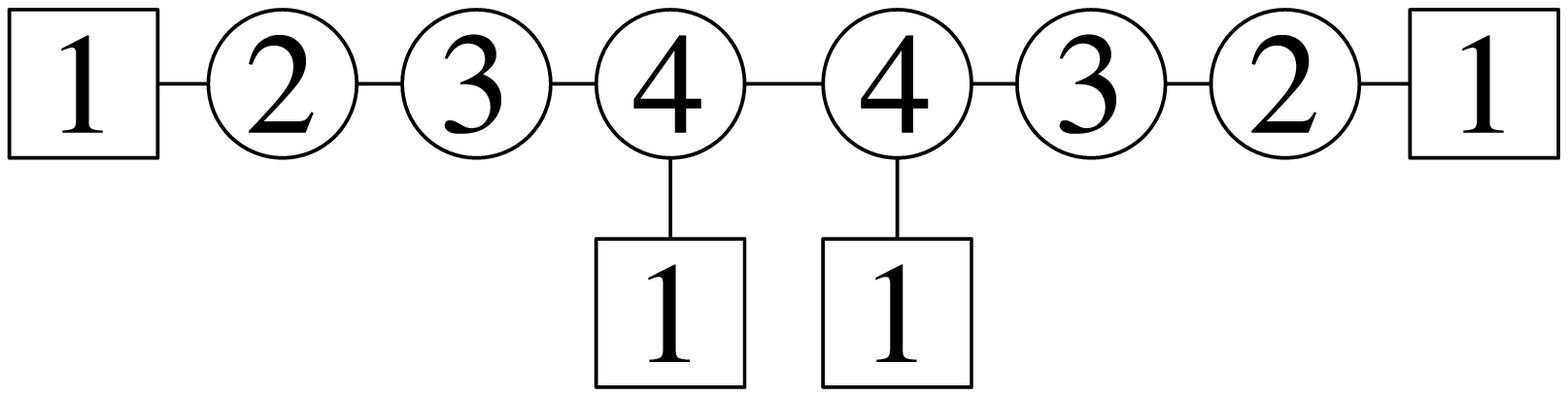} & \inc{\sizeA}{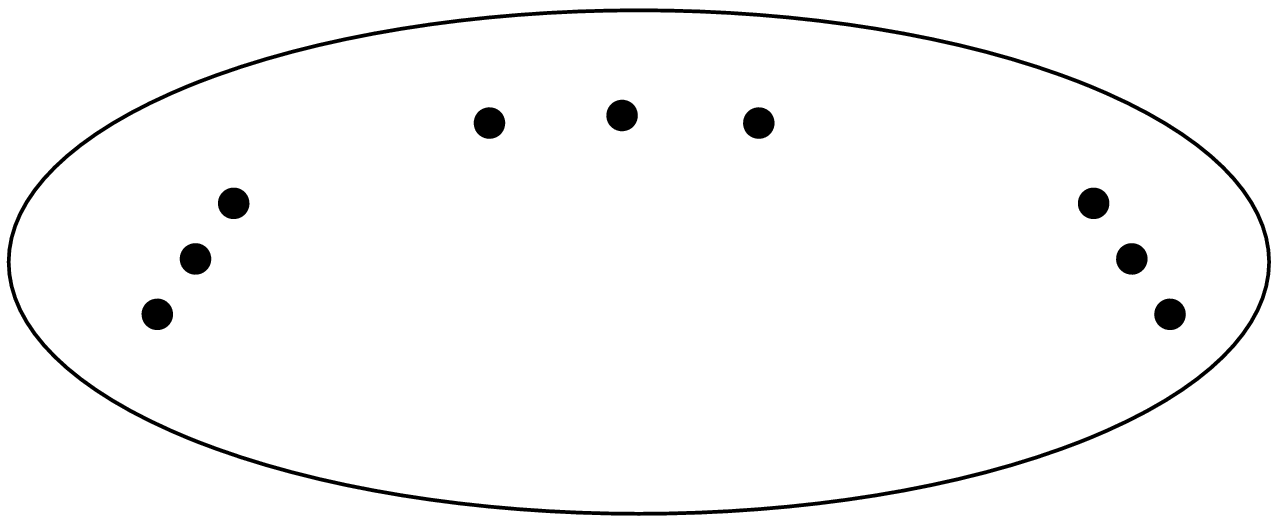} \\[2em]
\Downarrow  & \Downarrow \\[1em]
 \inc{\sizeA}{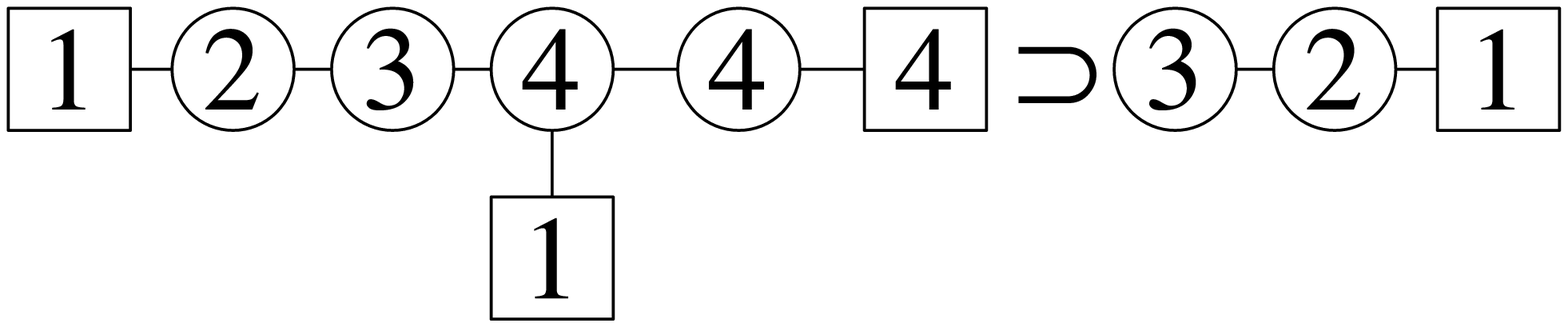} & \inc{\sizeA}{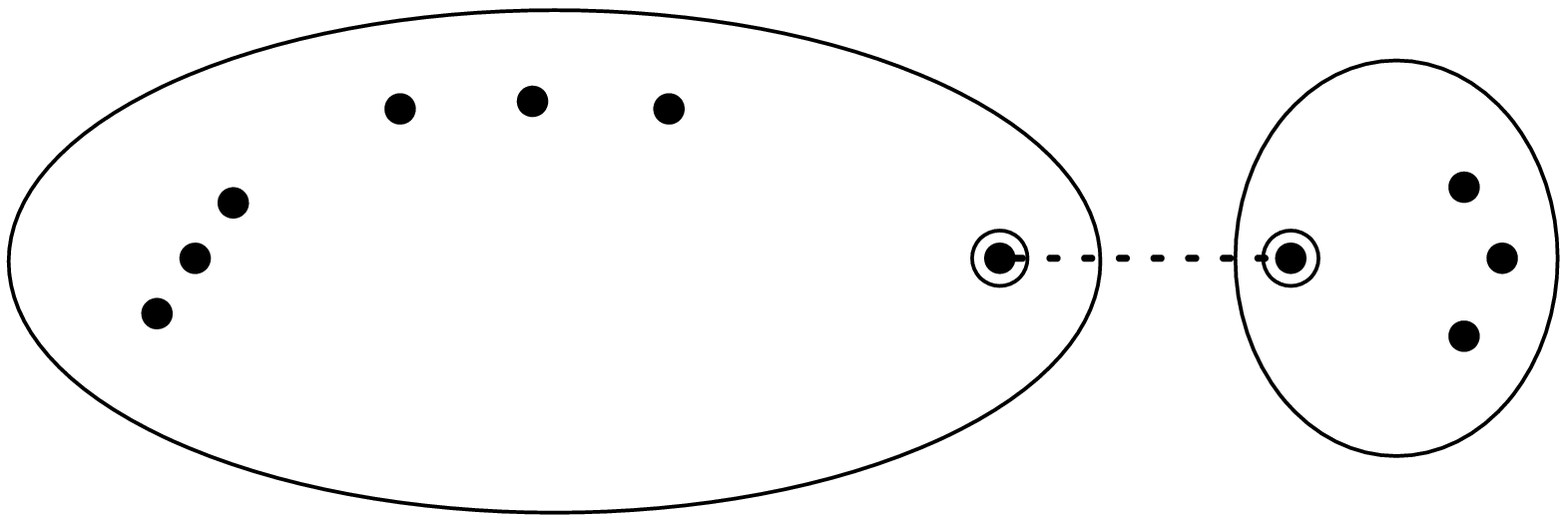} \\[2em]
\Downarrow  & \Downarrow \\[1em]
 \inc{\sizeA}{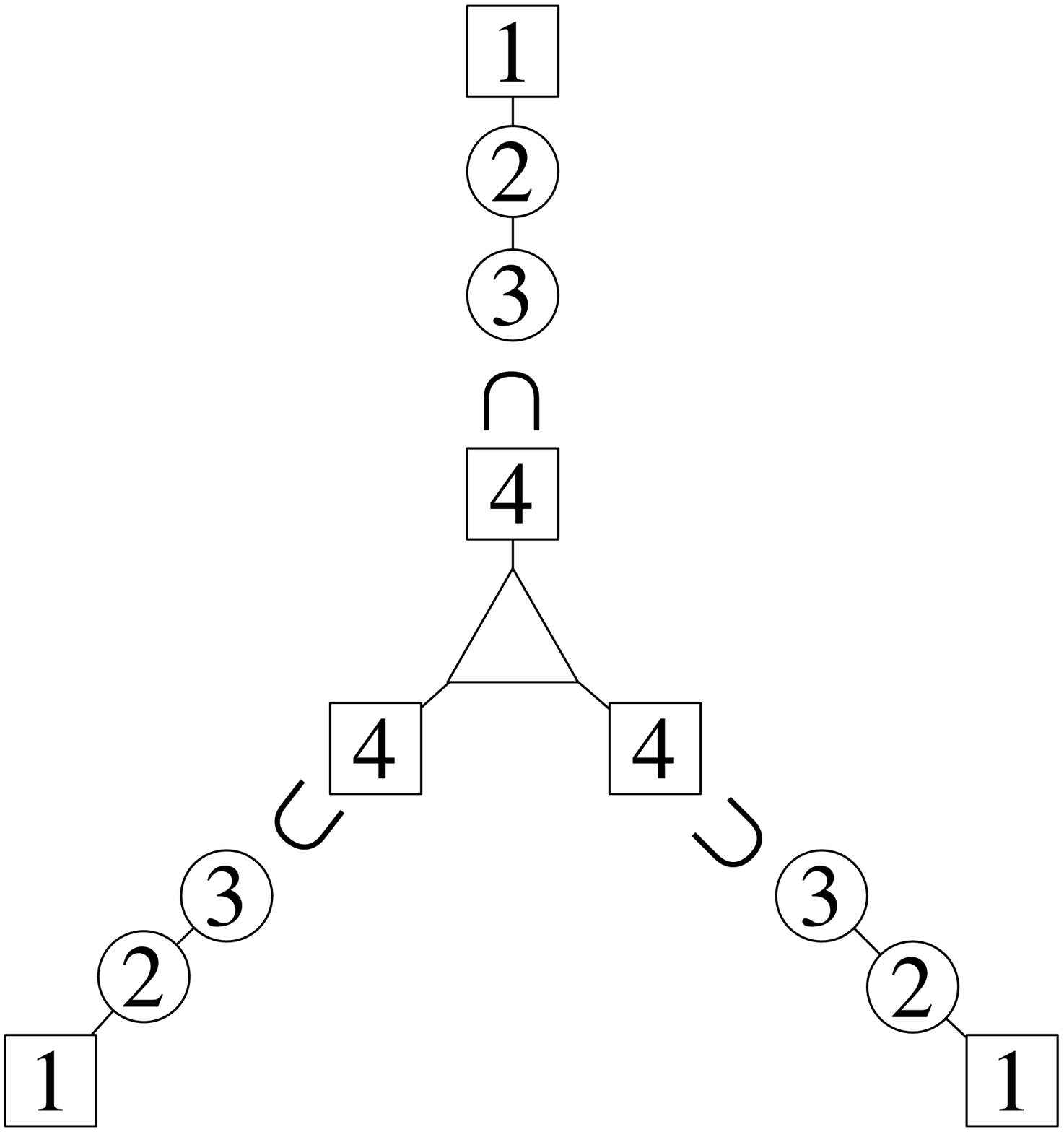} & \inc{\sizeA}{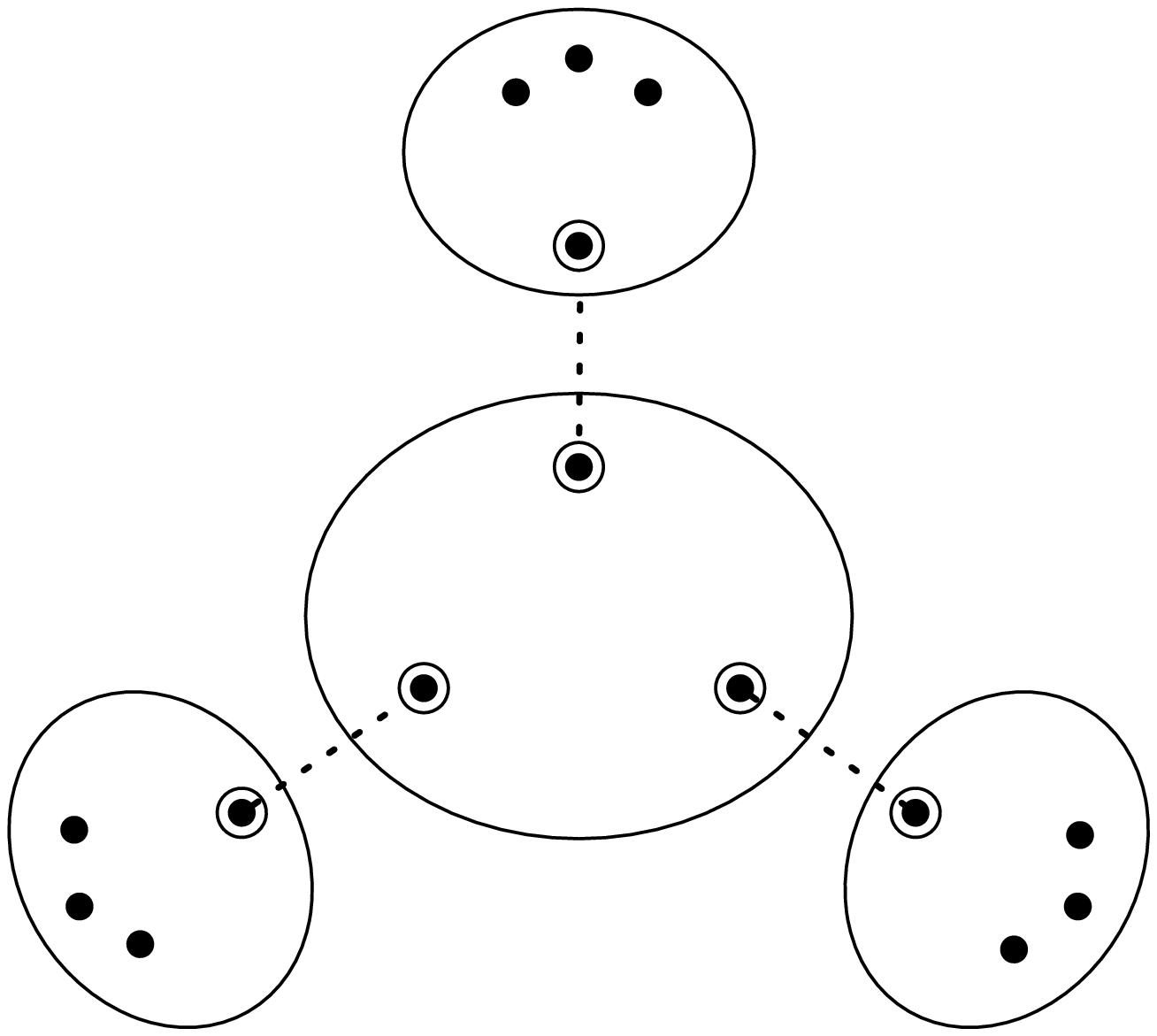} 
\end{array}
\]
\caption{Construction of $T_{\SU(4)}$. A circle or a box with $n$ inside
is an $\SU(n)$ gauge group or flavor symmetry, respectively.
A line connecting two objects is a  bifundamental hypermultiplet.
The symbol $\subset$ means that the subgroup of the flavor symmetry 
indicated couples to the corresponding gauge field. The triangle
with three $\SU(4)$ flavor symmetry attached stands for the $T_{\SU(4)}$ theory.\label{TSU4}}
\end{figure}
We start from the linear quiver shown in the first row of Fig.~\ref{TSU4},
and go to the region of the moduli space where three necks develop in the G-curve.
Originally one has a $\CP^1$ with nine punctures of type $\bullet$;
we split off three spheres, each with three simple punctures.
Each endpoint of the necks becomes a full puncture $\odot$.
Let us first split one sphere with three simple punctures,
see the second row of Fig.~\ref{TSU4}.
The $\SU(3)$ group in the tail \begin{equation}
\SU(3)\times \SU(2)
\end{equation} becomes weakly coupled, and gauges
the subgroup of the $\SU(4)$ flavor symmetry associated to the puncture. 
We repeat the process three times,
and arrive at the situation  shown in the third row of Fig.~\ref{TSU4}.
The resulting theory was called  $T[A_3]$ in \cite{Gaiotto:2009we}
and  $T_4$ in \cite{Gaiotto:2009gz}.
We call it $T_{\SU(4)}$ to reduce possible later confusion.

$T_{\SU(4)}$ does not have a marginal coupling constant, because
a configuration consisting of three points on a sphere has no modulus. 
The pole structure at each of the punctures is that $\phi_{2,3,4}$ has poles of order $1,2,3$.
Applying the formula \eqref{counting}, one concludes that $T_{\SU(4)}$ 
has one operator of dimension $3$
and two operators of dimension $4$.
Then one can check that the quivers shown in Fig.~\ref{TSU4} 
have the same number
of Coulomb branch operators for each scaling dimension.

The central charges $a$ and $c$ of this theory
can also be easily calculated because they are independent
of exactly marginal deformations. 
It is more intuitive to parametrize the central charges $a$ and $c$
using the effective number of hyper- and vector multiplets $n_v$ and $n_h$,
as defined by the relation:
\begin{equation}
a=\frac{5n_v+n_h}{24},\qquad
c=\frac{2n_v+n_h}{12}.
\end{equation} 
$n_v$ and $n_h$ of the total theory are
obtained by counting the number of multiplets at the original weakly-coupled limit:
\begin{equation}
n_v(\text{total})=52, \qquad n_h(\text{total})=64.
\end{equation} In the regime where the three necks develop, one has three tails, each with
\begin{equation}
n_v(\text{tail})=11, \qquad n_h(\text{tail})=8,
\end{equation} Therefore we have \begin{equation}
n_v(T_{\SU(4)}) = 52-3\times 11=19,\qquad
n_h(T_{\SU(4)}) = 64 - 3\times 8 = 40. \label{acOfTSU4}
\end{equation}

\section{ 6d $D_{N}$ theory and $\SO$--$\USp$ quivers} \label{Dtype}
\subsection{Preliminary comments on $\SO$ and $\USp$ gauge groups}\label{SOpre}
Having reviewed the construction the $\SU$ quivers, here we start the analysis
of the $\SO$--$\USp$ quivers.
First we need to recall rudiments of these gauge groups, and also a few properties
of hypermultiplets.

What is usually called a hypermultiplet in the representation $R$ of a group $G$
consists of an $\cN=1$ chiral multiplet in the representation $R$ and another
in the conjugate representation $R^*$. When we have $N$ copies of them
the flavor symmetry is at least $\U(N)$. When $R$ is strictly real, it enhances
to $\USp(2N)$, as can be understood from the form of the $\cN=1$ superpotential.
When $R$ is pseudo-real, one $\cN=1$ chiral multiplet in $R$ forms an
$\cN=2$ hypermultiplet, which is called a {\em half-hypermultiplet} in $R$.
When we have $N$ copies of them, the flavor symmetry is $\SO(N)$.

Now let us consider an $\SO(n)$ gauge theory with $N_f$ massless hypermultiplets
in the vector representation of dimension $n$.
It has $\USp(2N_f)$ flavor symmetry because the vector representation is strictly real.
Here and in the following we use the convention that the fundamental
representation of $\USp(2n)$ is of dimension $2n$; thus in our notation 
$\USp(2)\simeq \SU(2)$ and $\USp(4)\simeq \SO(5)$ at the level of Lie algebra.
The gauge coupling constant is marginal when $N_f=n-2$.

Next consider a $\USp(2n)$ gauge theory with $N_f$ hypermultiplets
in the vector representation.
The flavor symmetry is then $\SO(2N_f)$, and 
the theory becomes superconformal when $N_f=2n+2$.
It will be important that the vector representation,
which is $2n$-dimensional, is pseudo-real.
This implies that one can form half-hypermultiplets, although
one cannot have an odd number of half-hypermultiplets because of Witten's global anomaly.
One can still gauge the subgroup $\SO(d)\times \SO(2N_f-d)\subset \SO(2N_f)$
for odd $d$, preserving $\cN=2$ supersymmetry.

Therefore one can naturally consider a quiver theory
with alternating gauge groups \begin{equation}
\cdots \times \SO(d_a)\times \USp(d_{a+1}-2)\times \SO(d_{a+2})\times 
\USp(d_{a+3}-2) \times \cdots
\end{equation} with bifundamental half-hypermultiplets between
consecutive gauge groups, and possibly with extra hypermultiplets
in the fundamental representation for the $a$-th gauge group.
Here the bifundamental representation is the tensor product of the vector
representation of $\SO$ group and the fundamental representation of $\USp$ group.
We let $k_a$ be  twice the number of hypermultiplets in the vector representation
if the $a$-th gauge group is $\SO$, while
we let it be the number of half-hypermultiplets in the fundamental representation
if the $a$-th gauge group is $\USp$.
Then the flavor symmetry is $\USp(k_a)$ and $\SO(k_a)$, respectively.
For convenience we define $\delta_a=0$ when the $a$-th gauge group is $\SO$,
and $\delta_a=2$ when it is $\USp$.

\begin{figure}
\centerline{
	\inc{\sizeA}{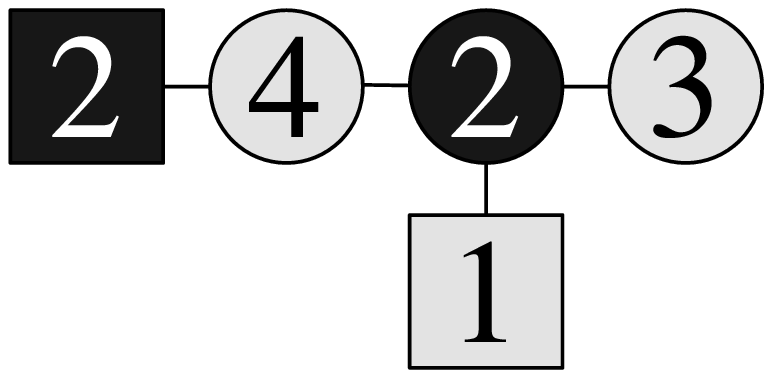} \qquad
	\inc{\sizeC}{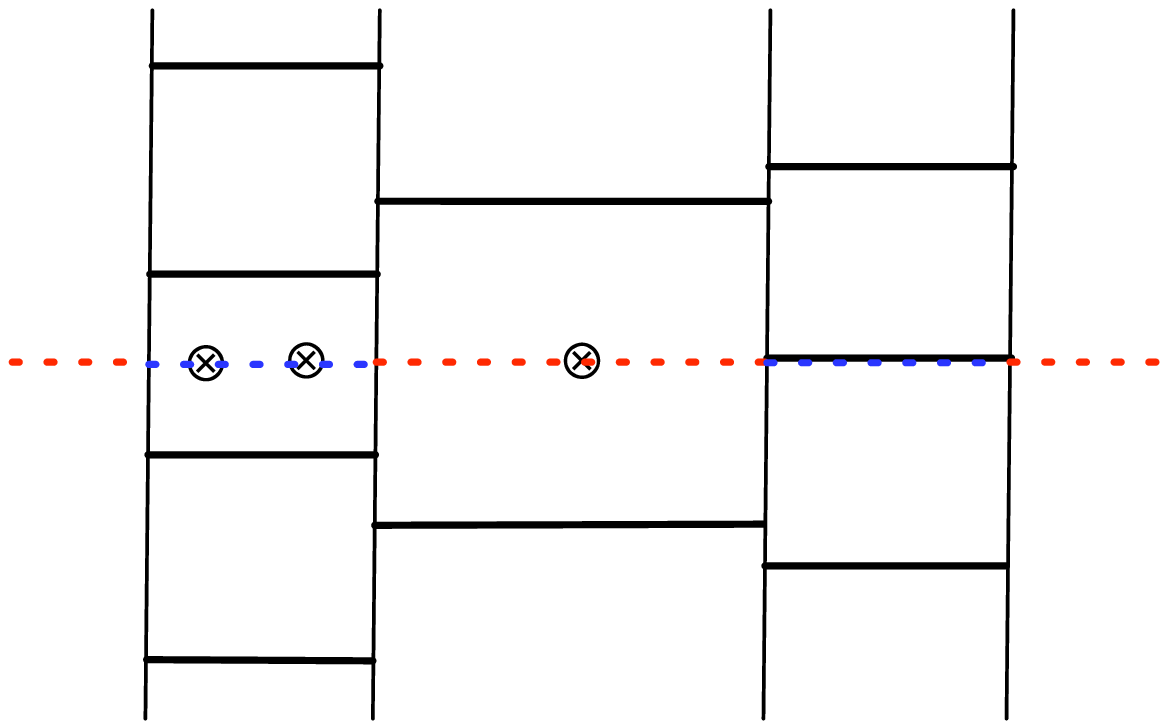}
}
\caption{On the left:
an example of $\SO$--$\USp$ quiver gauge theory.
A circle or a box stands for a gauge group or a flavor symmetry, respectively.
A gray object with $n$ inside is an $\SO(n)$ group,
a black object with $n$ inside is a $\USp(n)$ group.
On the right: the brane configuration realizing the quiver.
The vertical lines stand for NS5-branes,
the horizontal lines D4-branes suspended between them,
and $\otimes$ D6 branes. The dotted line represents the O4-plane.
The color distinguishes two types of O4-planes. \label{braneconfig}}
\end{figure}

The requirement of marginality of each of the gauge coupling constants
can be written succinctly as \begin{equation}
k_a=2d_a-d_{a-1}-d_{a+1},
\end{equation} exactly as in the case of the quiver of $\SU$ gauge groups \eqref{SUmarginalitycondition}.
One such superconformal quiver is shown in Fig~\ref{braneconfig}.
There, a box stands for a flavor symmetry, and a circle a gauge symmetry;
a gray one with  $n$ inside is an $\SO(n)$ group,
and a black one is an $\USp(n)$ group; a line stands for a half-hypermultiplet
in the bifundamental representation. The theory shown  
thus has the gauge group \begin{equation}
\SO(4)\times\USp(2)\times \SO(3)
\end{equation} with bifundamentals,
one extra hypermultiplet in the vector representation of $\SO(4)$,
and one extra half-hypermultiplet in the fundamental representation of $\USp(2)$.

\subsection{From type IIA brane configuration to the G-curve}\label{brane}
Let us realize the quiver theory introduced in the previous subsection via
a system of NS5, D4 and D6 branes with orientifolds in type IIA string theory,
which is schematically shown in Fig.~\ref{braneconfig}.
These systems and the corresponding Seiberg-Witten curves were first analyzed by 
\cite{Evans:1997hk,Landsteiner:1997vd,Brandhuber:1997cc}; the subtler aspects of the orientifolding procedure
was later clarified in  \cite{deBoer:1998by,Hori:1998iv,Gimon:1998be}.

We start from the flat ten-dimensional spacetime,
and put $n+1$ NS5 branes extending along directions $x^{0,1,2,3,4,5}$.
We perform the  orientifolding, flipping directions $x^{4,5,7,8,9}$,
which introduces an O4-plane in the system.
One important aspect is that an O4$^-$-plane becomes an O4$^+$-plane
when it crosses an  NS5-brane, and vice versa.
We define $\delta_a$  to be $0$ or $2$ depending on the type of the O4-plane
between $a$-th and $(a+1)$-st NS5 brane,
so that we have an $\SO$ gauge group when $\delta_a=0$
and a $\USp$ gauge group when $\delta_a=2$.
We analogously define 
$\delta_0$ and $\delta_{n+1}$.
$\delta_a$ accounts the difference of D4-charge carried by an O4$^-$-plane
and an O4$^+$ plane.

We then suspend $d_a-\delta_a$ 
D4-branes between the $a$-th and $(a+1)$-th NS5-branes.
The $a$-th gauge group is $G_a=\SO(d_a)$ if $\delta_a=0$
and $=\USp(d_a-\delta_a)$ if $\delta_a=2$.
We denote by $k_a$
the number of D6-branes, extending along $x^{0,1,2,3,7,8,9}$, 
between the $a$-th and $(a+1)$-th NS5-branes;
for simplicity we put all D6-branes on top of the O4-plane.
This configuration realizes in the low-energy limit
 the quiver gauge theory specified
by  the sequences of numbers $(d_a)$, $(\delta_a)$ and  $(k_a)$,
as discussed in the previous subsection.

One example is depicted in Fig.~\ref{braneconfig}:
There, a vertical line stands for a NS5-brane, a horizontal solid line a D4-brane,
a horizontal dotted line an O4-plane and an $\otimes$ a D6-brane.
An O4$^-$-plane is in blue and an O4$^+$-plane is in red.
We  have the gauge group $\SO(4)\times \USp(2)\times \SO(3)$
with bifundamentals between consecutive gauge factors,
and extra hypermultiplets  in the fundamental representations for $\SO(4)$ and $\USp(2)$.
The properties of the O4$^+$ plane to the left and to the right of
the D6-brane on top of it is known to be slightly different,
and the plane to the right is, properly speaking,
an $\widetilde{\text{O4}}{}^+$-plane, which is important to guarantee that there is no 
Witten's global anomaly in the low energy gauge theory, see \cite{Hori:1998iv} for details.

Let us now consider a quiver with the gauge groups \begin{equation}
\USp(2N-2)\times \SO(2N)\times \USp(2N-2) \times \cdots \times \SO(2N)\times \USp(2N-2),
\end{equation} with a total of $n=2s+1$ gauge factors,
and with $2N$ massless half-hypermultiplets in the fundamental representation
for each of the two $\USp(2N)$ gauge groups at the ends to make them superconformal.
The Seiberg-Witten curve is given by \cite{Landsteiner:1997vd}
\begin{equation}
F(v,t)=v^{2N}t^{n+1} + P_1(v^2)t^n+P_2(v^2)t^{n-1}+\cdots+P_n(v^2)t+v^{2N}=0,
\label{soSWcurve}
\end{equation}where \begin{equation}
P_i(v^2)=c_i v^{2N}  + u_i^{(2)}v^{2N-2} + u_i^{(4)}v^{2N-4} +\cdots + u_i^{(2N)}.
\end{equation} Here $u_i^{(2k)}$ is the Casimir of degree $2k$ of the $i$-th gauge group,
except for $u_i^{(2N)}$ when the $i$-th gauge group is $\USp(2N-2)$, for which
no such Casimir exists. In fact, the zeros of $F(v,t)$ at $v=0$ 
all need to be double zeros:
\begin{equation}
F(0,t)=u_1^{(2N)} t^n + u_2^{(2N)} t^{n-1} + \cdots + u_n^{(2N)} t 
=  \alpha t^2 \prod_{i=1}^{s-1} (t-q_i)^2\label{doublezero}
\end{equation} for some choice of $\alpha$ and  $q_i$.
In particular this forces $u_1^{(2N)}=u_n^{(2N)}=0$.
This condition leaves $s$ independent parameters $\alpha$ and $q_i$,
which encode the Casimir operators of $s$ $\SO(2N)$ gauge factors.

This condition is necessary to prevent a so-called ``$t$-configuration,''
i.e. a transversal intersection of a single M5-brane 
with the M-theory orientifold plane
\cite{deBoer:1998by,Hori:1998iv}. 
 
The Seiberg-Witten curve can be rewritten in Gaiotto's form \begin{equation}
0=x^{2N}+ \varphi_{2} x^{2N-2} + \varphi_{4} x^{2N-4} +\cdots + \varphi_{2N},
\end{equation} where $x= vdt/t$
is a holomorphic differential on the G-curve 
$\Sigma=\CP^1$ parametrized by $t$.
$\varphi_{2k}$ encodes the vevs of Coulomb branch operators:
\begin{equation}
\varphi_{2k}= 
\frac{u_1^{(2k)}t^{n}+u_2^{(2k)}t^{n-1}+\cdots+ u_n^{(2k)} t }
{\prod (t-t_a)}
\left(\frac{dt}{t}\right)^{2k}.\label{baz}
\end{equation}  
Here the $t_a$ are defined by  \begin{equation}
\prod(t-t_a) =t^{n+1} +c_1 t^{n} + \cdots + c_n t + 1\label{tDef}
\end{equation} which encodes the gauge coupling constants.
We see that 
$\varphi_{2k}$ is allowed to have a simple pole
at $n+1$ points where $t=t_i$,
whereas it is allowed to have poles of order $2k-1$
at two points  $t=0$ and $t=\infty$.

As for $\varphi_{2N}$, the condition \eqref{doublezero} means that it can be written as
\begin{equation}
\varphi_{2N}=\varphi_{\tilde N}^2 \quad\text{where}\quad
\varphi_{\tilde N}= \frac{t\prod (t-q_i)}{\prod (t-t_a)^{1/2}} (\frac{dt}{t})^N.
\end{equation} 
$\varphi_{\tilde N}$ has $\bZ_2$ monodromy around $t=t_a$, with a pole of order $1/2$;
whereas it has no monodromy around $t=0,\infty$ and has poles of order $N-1$.

\begin{figure}
\[
\begin{array}{c@{\quad}c@{\qquad}c}
1) & \inc{\sizeA}{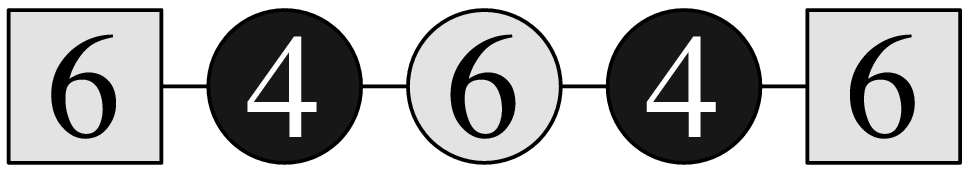} & \inc{\sizeA}{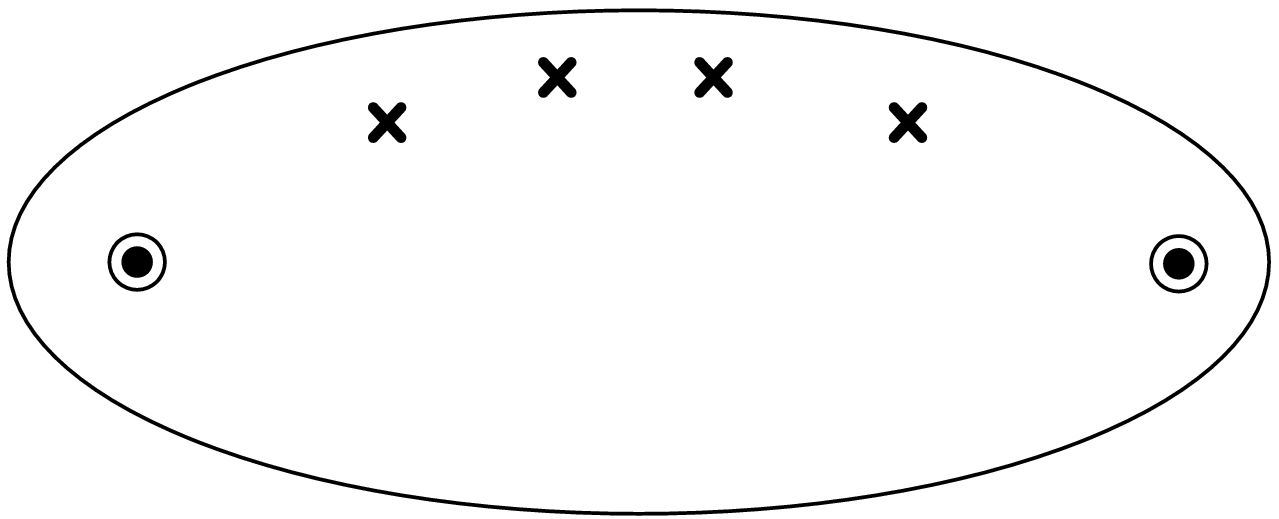} \\[2em]
2) & \inc{\sizeA}{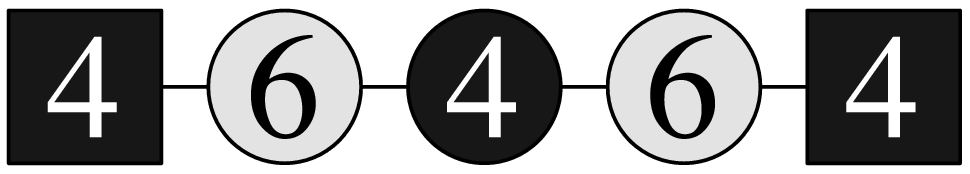} & \inc{\sizeA}{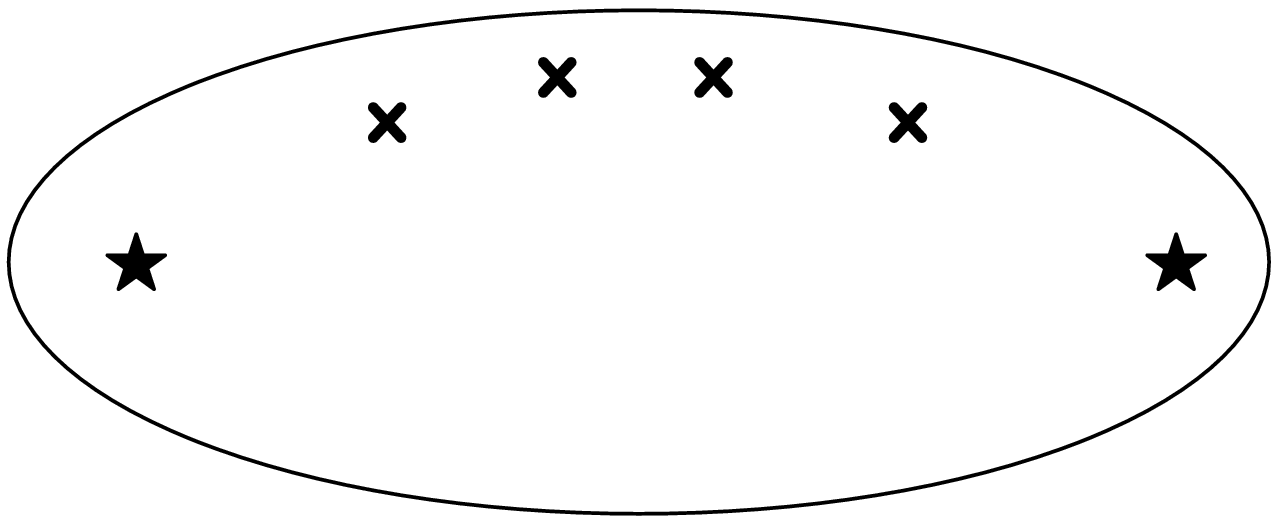} \\[2em]
3) & \inc{\sizeA}{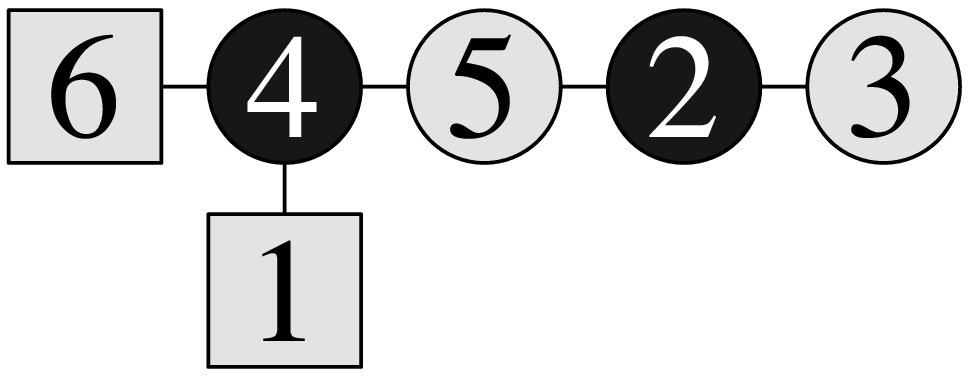} & \inc{\sizeA}{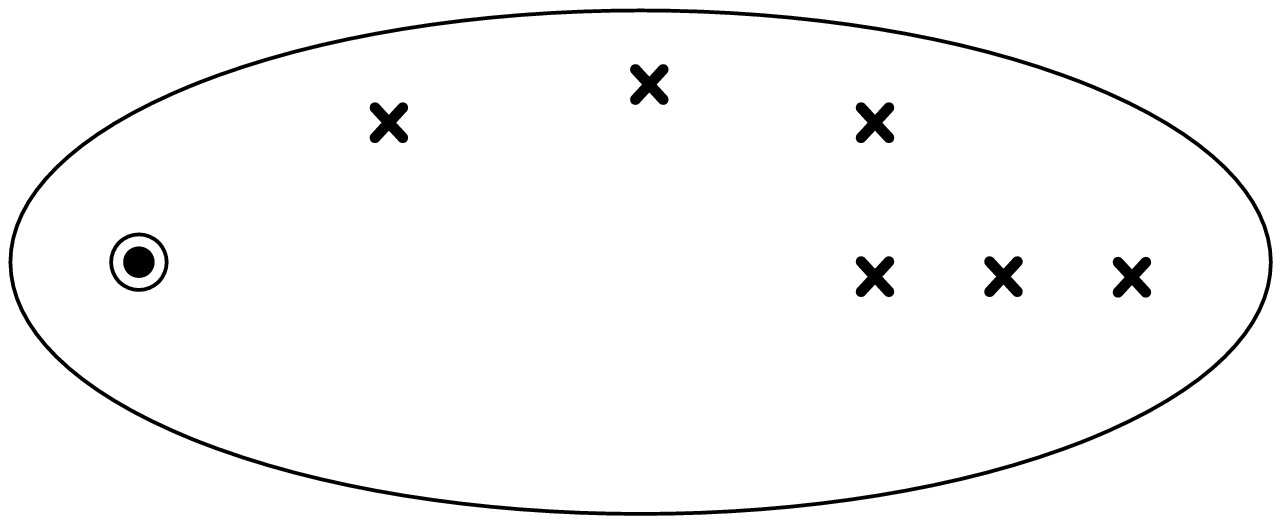}  
\end{array}
\]
\caption{Examples of $\SO$--$\USp$ quivers and their G-curves. 
Simple punctures are marked by $\times$.
There are two types of full punctures.
Each of the punctures labeled by $\odot$ or $\star$ has one $\SO(2N)$ or
$\USp(2N-2)$ flavor symmetry, respectively.
 \label{so6example}}
\end{figure}

The G-curve is shown as the first example in  Fig.~\ref{so6example};
there, we have taken $2N=6$ and $n=3$. 
$\varphi_{2k}$ has simple poles and $\varphi_{\tilde N}$ behaves as $\sim 1/t^{1/2}$
at the punctures denoted by $\times$ with the local coordinate $t$ chosen such that the puncture
is at $t=0$.
$\varphi_{2k}$ has poles of order $2k-1$ and $\varphi_{\tilde N}$ has poles of order $N-1$
at the punctures denoted by $\odot$.

Again, this system can be thought of as the compactification of
the six-dimensional $D_N$ theory on $\Sigma$, with prescribed sets of singularities
for the worldvolume fields.
Recall that the six-dimensional $D_N$ theory arises as the low-energy 
theory on a stack of $2N$ M5-branes on top of the $\bR^5/\bZ^2$
M-theory orientifold, or equivalently of the compactification of type IIB string theory
on an ALE space of type $D_N$.
This theory has operators of dimension $2,4,\ldots,2N-2$ and
one extra operator of dimension $N$, which become the differentials
$\varphi_{2}$, $\varphi_4$, \ldots, $\varphi_{2N-2}$ 
and $\varphi_{\tilde N}$, respectively.
The Lie algebra  of type $D_N$ has one outer automorphism,
under which operators of dimension $2,4,\ldots,2N-2$ are even
but the extra one of dimension $N$  is odd; it is the Pfaffian.
The analysis above shows that the simple puncture $\times$ has
a $\bZ_2$ monodromy of this outer automorphism associated to it.
This agrees with the known fact that the transversal intersection of an M5-brane
with the M-theory orientifold $\bR^5/\bZ_2$  screened by an even number of M5-branes
has an associated $\bZ_2$ charge \cite{Hori:1998iv}.

Let us next consider the quiver with the gauge groups \begin{equation}
\SO(2N)\times \USp(2N-2) \times \cdots \times \USp(2N-2)\times \SO(2N),
\end{equation} with a total of $2s+1$ gauge groups.
There are bifundamental hypermultiplets as always,
and $N-1$ hypermultiplets in the fundamental representation
for each of the $\SO(N)$ groups at the ends.
The Seiberg-Witten curve is given again by \eqref{soSWcurve},
but the condition on the double zeros is now \begin{equation}
F(0,t)=u_1^{(2N)} t^n + u_2^{(2N)} t^{n-1} + \cdots + u_n^{(2N)} t 
=  \alpha  t \prod_{i=1}^{s} (t-q_i)^2.
\end{equation}
There are $s$ simple punctures at $t_i$  as before,
but the pole structure at $t=0,\infty$ is now different:
$\varphi_{2k}$ still has  poles of order $2k-1$,
but $\varphi_{\tilde N}$ has a pole of order $N-1/2$.
In particular there is a $\bZ_2$ monodromy around $t=0,\infty$.
We label this type of punctures by $\star$.
The case $2N=6$ is shown in the second line of Fig.~\ref{so6example}.

The same exercise can be repeated with the quiver of the form \begin{equation}
\USp(2N-2)\times \SO(2N-1) \times \USp(2N-4) \times \SO(2N-3) \times
\cdots
\times \USp(2)\times \SO(3),
\end{equation} with bifundamentals between each pair of two consecutive gauge groups
as always, and $2N+1$ extra fundamental half-hypermultiplets
on the leftmost $\USp(2N-2)$ gauge group.
This quiver theory has a total of $2N-2$ gauge groups, and we find that the
resulting G-curve has $2N$ punctures of type $\times$
and one puncture of type $\odot$; see Appendix~\ref{curves} for details.
The case $2N=6$ is shown 
as the third example in Fig.~\ref{so6example}.

\subsection{SCFT with $\SO(2N)^3$ flavor symmetry}\label{SO6pants}

By making use of 
the interpretation of $\SO$--$\USp$ quivers
as  compactifications of the six-dimensional $D_N$ theory,
one can easily find their various infinitely strongly-coupled limits.
As an exercise let us construct a theory with no marginal coupling constant
and with $\SO(2N)^3$ flavor symmetry, which we denote as $T_{\SO(2N)}$.
The construction for the $A_N$ theory was 
reviewed in Sec.~\ref{SUpants}, which we closely follow.

\begin{figure}
\[
\begin{array}{c@{\qquad}c}
\inc{\sizeA}{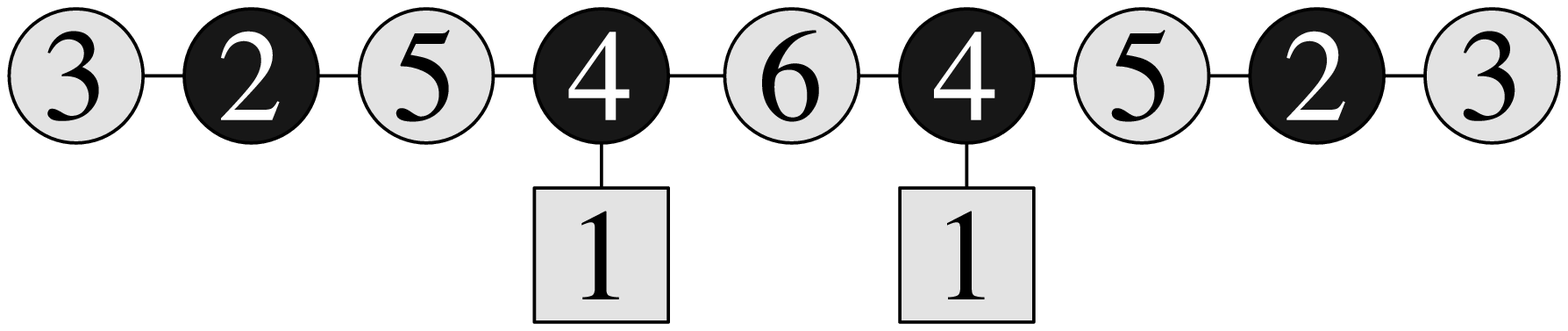} & \inc{\sizeA}{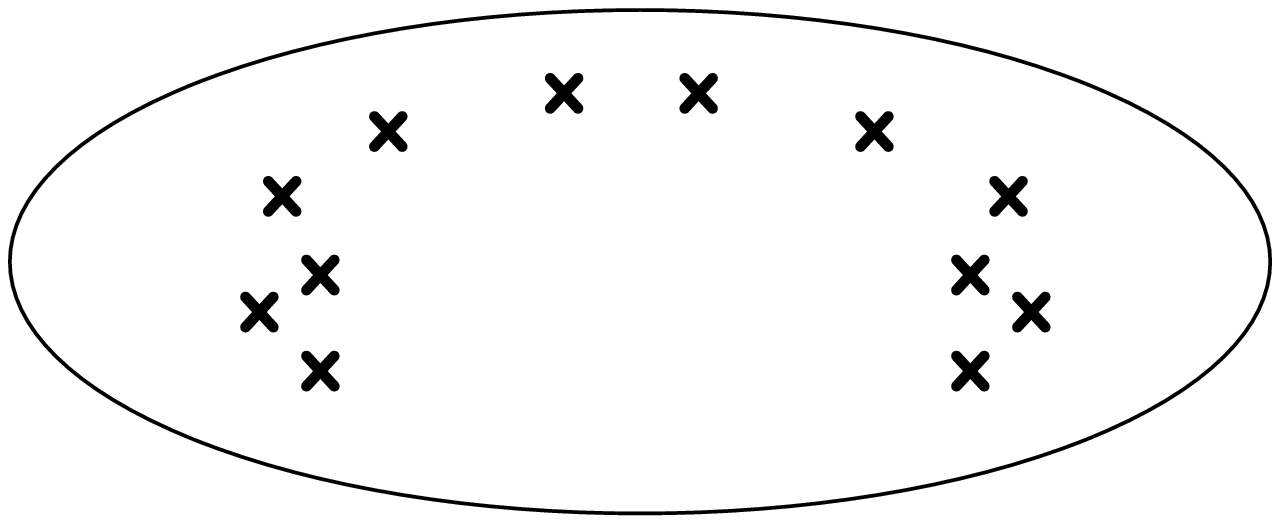} \\[2em]
\Downarrow  & \Downarrow \\[1em]
\inc{\sizeA}{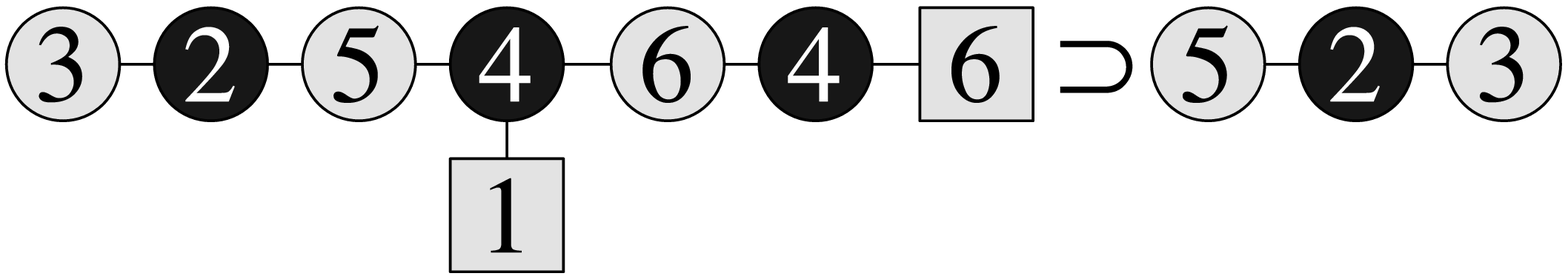} & \inc{\sizeA}{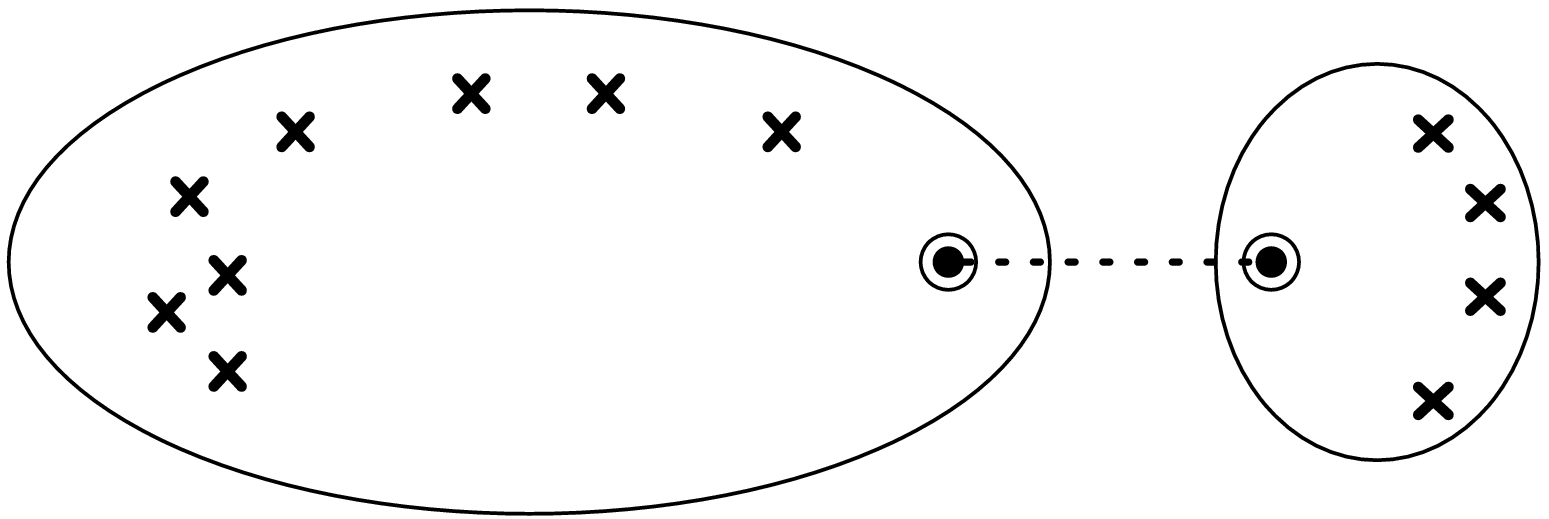} \\[2em]
\Downarrow  & \Downarrow \\[1em]
\inc{\sizeA}{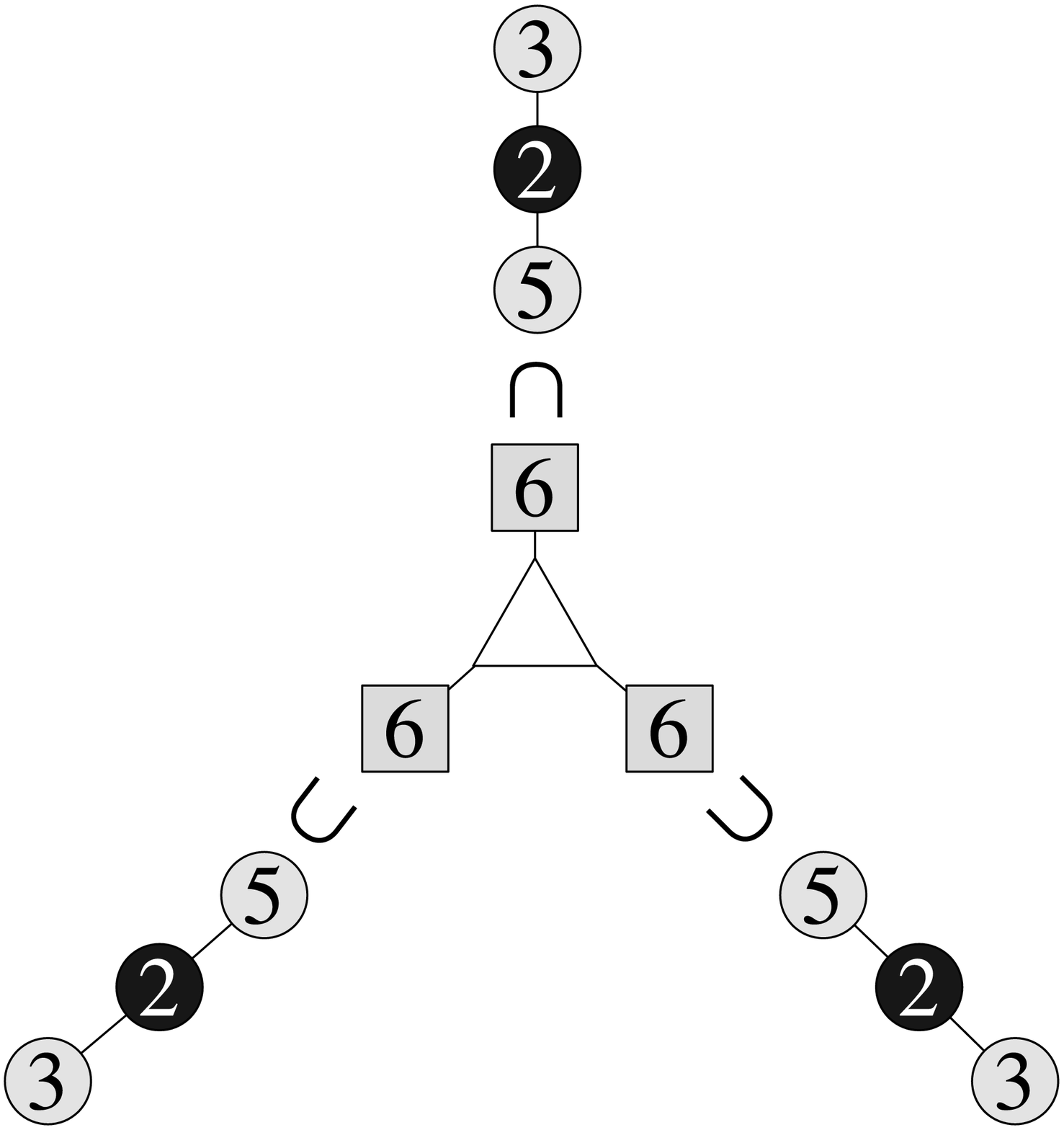} & \inc{\sizeA}{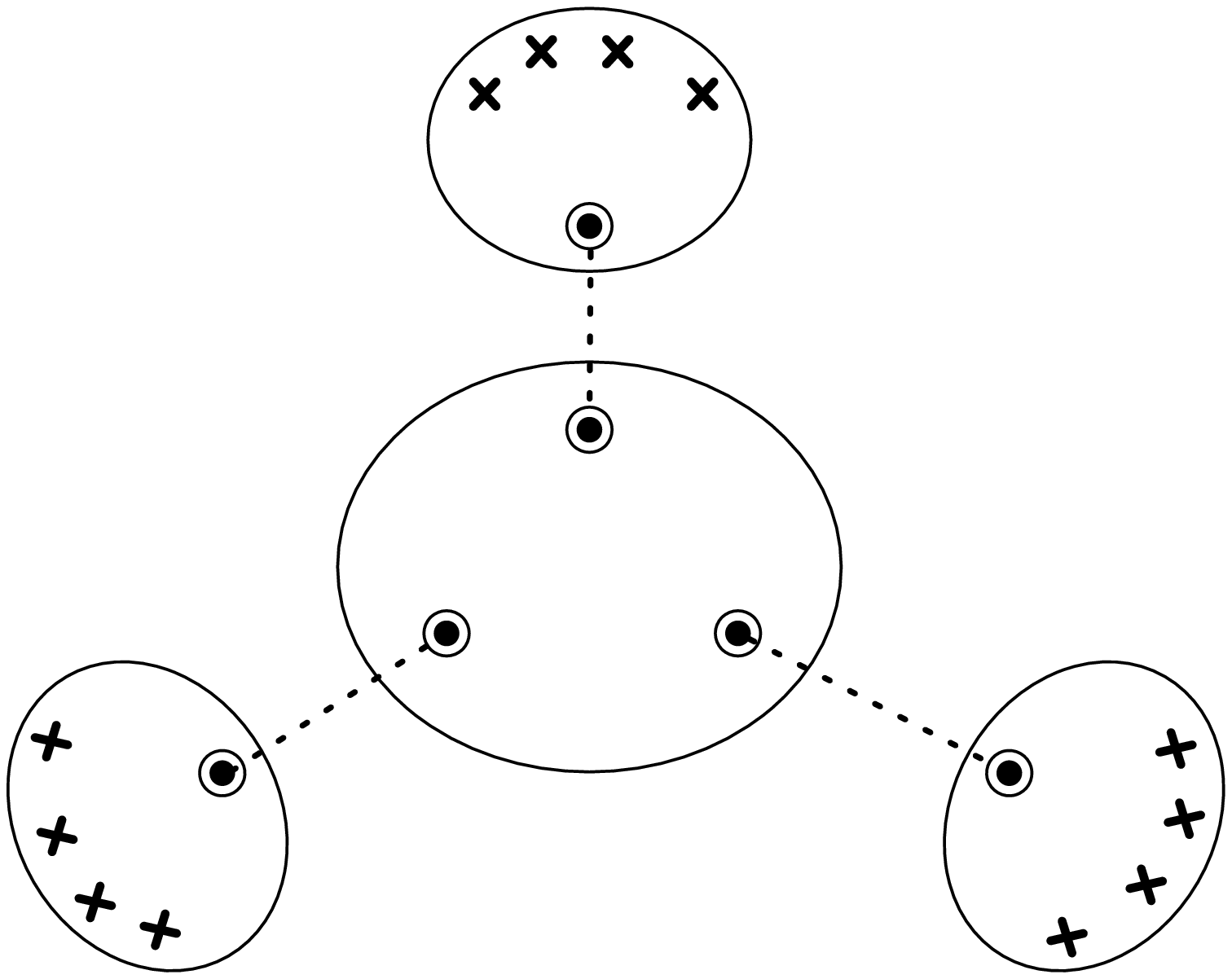} 
\end{array}
\]
\caption{Construction of $T_{\SO(6)}$. The triangle with 
three $\SO(6)$ flavor symmetries attached stands for the $T_{\SO(6)}$.
The symbol $\subset$  between the $\SO(6)$ flavor symmetry and the $\SO(5)$ gauge symmetry
signifies that $\SO(5)\subset \SO(6)$ is gauged.
 \label{TSO6}}
\end{figure}

For concreteness, let us first consider the case $2N=6$.
Take two copies of the quiver theory of the third example of Fig.~\ref{so6example},
and introduce an $\SO(6)$ gauge group which gauges the $\SO(6)$
flavor symmetry of the original one. 
One then has the linear quiver shown in the first row of Fig.~\ref{TSO6},
whose associated G-curve is a $\CP^1$ with twelve punctures of  type $\times$.
Let us go to the region of the moduli space where three necks develop.
We split off three spheres, each with four punctures of type $\times$.
Each of the endpoints of  the necks becomes a  full puncture marked by $\odot$.

Let us first split off one sphere with three simple punctures,
see the second row of Fig.~\ref{TSO6}. The group $\SO(5)$ in the tail
\begin{equation}
\SO(5)\times \USp(2)\times \SO(3)
\end{equation}  becomes weakly coupled, and gauges
the subgroup of the $\SO(6)$ flavor symmetry associated to the puncture. 
We repeat the process three times,
and arrive at the situation  shown in the third row of Fig.~\ref{TSO6}.
Again, $n_v$ and $n_h$ of the total theory 
are obtained by counting the number of multiplets: first, the total theory has
\begin{equation}
n_v(\text{total})=67, \qquad n_h(\text{total})=64.
\end{equation} In the regime where the three necks develop, one has three tails, each with
\begin{equation}
n_v(\text{tail})=16, \qquad n_h(\text{tail})=8.
\end{equation} Therefore we have \begin{equation}
n_v(T_{\SO(6)}) = 67-3\times 16=19,\qquad
n_h(T_{\SO(6)}) = 64 - 3\times 8 = 40.
\end{equation}

These numbers are exactly the same as those for $T_{\SU(4)}$, \eqref{acOfTSU4}.
Recall that the $\SU(4)$ quivers arose from the compactification
of the worldvolume theory on four coincident M5 branes,
whereas the $\SO(6)$--$\USp(4)$ quivers arose from 
six coincident M5 branes on top of the $\bR^5/\bZ_2$ orientifold singularity.
These two systems give the same low-energy six-dimensional 
$A_3\simeq D_3$ $(2,0)$ theory.  Therefore, 
they should result in the same SCFT in four dimensions,
because we compactified the same theory on the same surface,
with the same number of the same type of defects.
The agreement of $n_v$ and $n_h$ of $T_{\SU(4)}$ and $T_{\SO(6)}$ is
expected from the six-dimensional  viewpoint, but  is quite nontrivial
from the perspective of four-dimensional gauge theory.

The construction of $T_{\SO(2N)}$ can be done analogously.
We start from a linear quiver with $6N-9$ gauge groups \begin{multline}
\SO(3)\times \USp(2)\times \cdots \USp(2N-4)\times \SO(2N-1) \times \\
\USp(2N-2) \times \SO(2N)\times \cdots \times \SO(2N) \times \USp(2N-2) \times \\
\SO(2n-1) \times \USp(2N-4) \times \cdots \times \USp(2) \times \SO(3),
\end{multline}
with a bifundamental half-hypermultiplet between each pair of two consecutive groups,
and one half-hypermultiplet in the fundamental for the first and the last $\USp(2N-2)$ 
gauge groups.
The G-curve then is a sphere with $3(2N-2)$ punctures of type $\times$.
One can split off three spheres with $2N-2$ punctures each,
thus decoupling three tails of the form \begin{equation}
\SO(3)\times \USp(2)\times \cdots \USp(2N-4)\times \SO(2N-1).
\end{equation}
This results in a theory described by a G-curve with three punctures of type $\odot$.
We then have \begin{equation}
n_v(T_{\SO(2N)})=\frac{8N^3}{3}-7N^2+\frac{10N}{3},\qquad
n_h(T_{\SO(2N)})=\frac{8N^3}{3}-4N^2+\frac{4N}{3}.
\end{equation}

Now we can paste multiple copies of $T_{\SO(2N)}$ by gauging
$\SO(2N)$ groups to find a four-dimensional realization of the compactification
of the six-dimensional $D_N$ theory. It would be interesting
to extend the holographic analysis of \cite{Gaiotto:2009gz} to this case 
and reproduce $n_h$ and $n_v$ from the gravity solution.

It is natural to ask if there is a theory whose G-curve is a sphere
with three punctures of type $\star$ and with 
$\USp(2N-2)^3$ flavor symmetry. 
This is impossible because 
one cannot have $\bZ_2$ monodromy at three points on the sphere.
Instead it is possible to construct a theory whose G-curve is a sphere
with two punctures of type $\star$ and one puncture of type $\odot$,
by performing a similar procedure to the one presented above.
The flavor symmetry is then $\SO(2N)\times \USp(2N-2)^2$.
The important point is that 
two punctures of type $\star$
appear when a G-curve degenerates and 
develops a neck with $\bZ_2$ monodromy around it.

\subsection{Tails, tableaux and flavor symmetries}
Let us now classify possible types of superconformal tails of the $\SO$--$\USp$ quivers.
We found in Sec.~\ref{SOpre} that the requirement of the marginality of coupling constants
implies  \begin{equation}
d_1 < d_2 < \cdots < d_l=d_{l+1}=\cdots=d_r > d_{r+1} > \cdots d_n.
\end{equation} 
We let $2N=d_l=\cdots = d_r$.
Then we can associate a Young tableau 
with rows of widths $d_r-d_{r+1}$, $d_{r+1}-d_{r+2}$,\ldots,
as was the case for the tails of $\SU$ quivers.
There are two crucial differences, however.
One is that we need to distinguish the cases 
for which the last gauge group is $\SO$ or $\USp$;
the other is that  not all of the tableaux are allowed because $d_a$ for 
a $\USp$  gauge group needs to be even.

For a given tail,  let us then associate a tableau with the following rule:
\begin{itemize}
\item If it ends with a $\USp$ group, associate a tableau, with gray boxes,
whose rows are of width $d_r-d_{r+1}$, $d_{r+1}-d_{r+2}$, \ldots. One has $2N$ boxes in total.
\item If it ends with an $\SO$ group, associate a tableau, with black boxes,
whose rows are again of width $d_{r}-d_{r+1}$, $d_{r+1}-d_{r+2}$, \ldots, {\em except} 
the last row, for which we let the width be $d_{n}-d_{n+1}-2 = d_n-2$. This procedure
is consistent because the smallest $\SO$ group one can have is $\SO(3)$. 
One has $2N-2$ boxes in total.
\end{itemize}
To help grasp the procedure,
we list all the tails of $\SO(6)$--$\USp(4)$ quivers in Table~\ref{SOtails} and in Table~\ref{USptails}.
Note that a tableau with one row of $2N$ gray boxes corresponds to the puncture
of type $\odot$, the tableau with one column of $2N-2$ black boxes to
the puncture of type $\times$,
and the tableau with one row of $2N-2$ black boxes to
the puncture of type $\star$
that we used in the previous subsection;
we use these notations interchangeably.

\begin{table}
\[
\begin{array}{@{\extracolsep{1ex}}c@{\hskip1ex}|c@{\hskip1ex}|c@{\hskip1ex}||c@{\hskip1ex}|c}
\text{Tableau} &  \text{Alias} &\text{Flavor} & \text{Quiver} & \text{G-curve} \\
\hline
\inc{\sizeB}{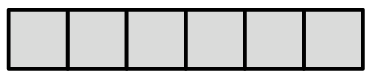} &\odot &  \SO(6) & \inc{\sizeA}{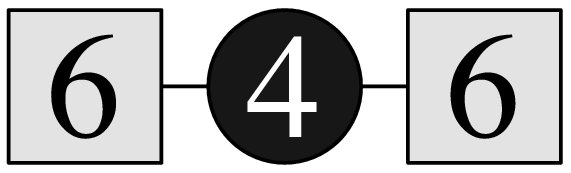}  & \hbox{\vbox to 2em{}}\inc{\sizeA}{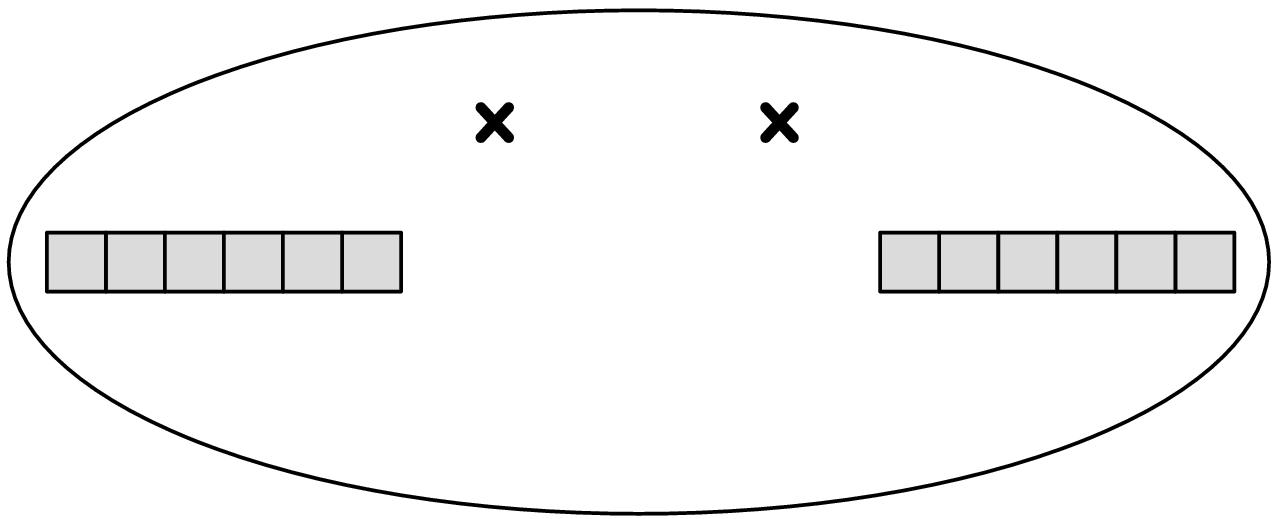}  \\[2em]
\inc{\sizeB}{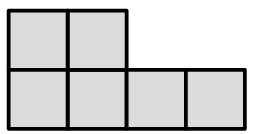} && \USp(2)\times \SO(2) & \inc{\sizeA}{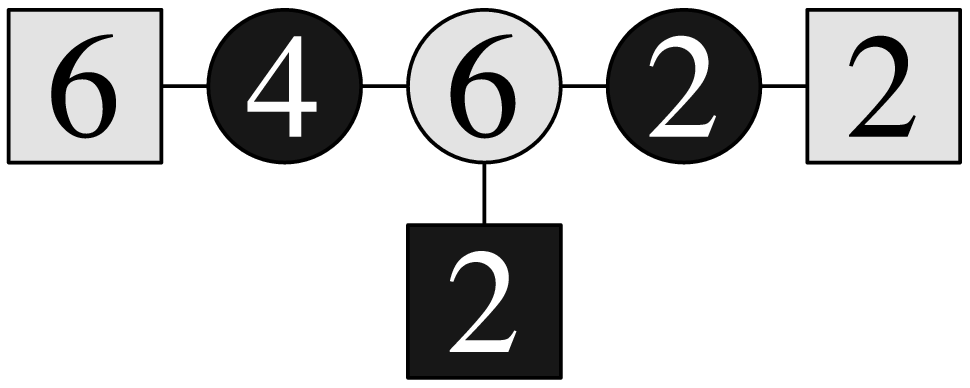} & \inc{\sizeA}{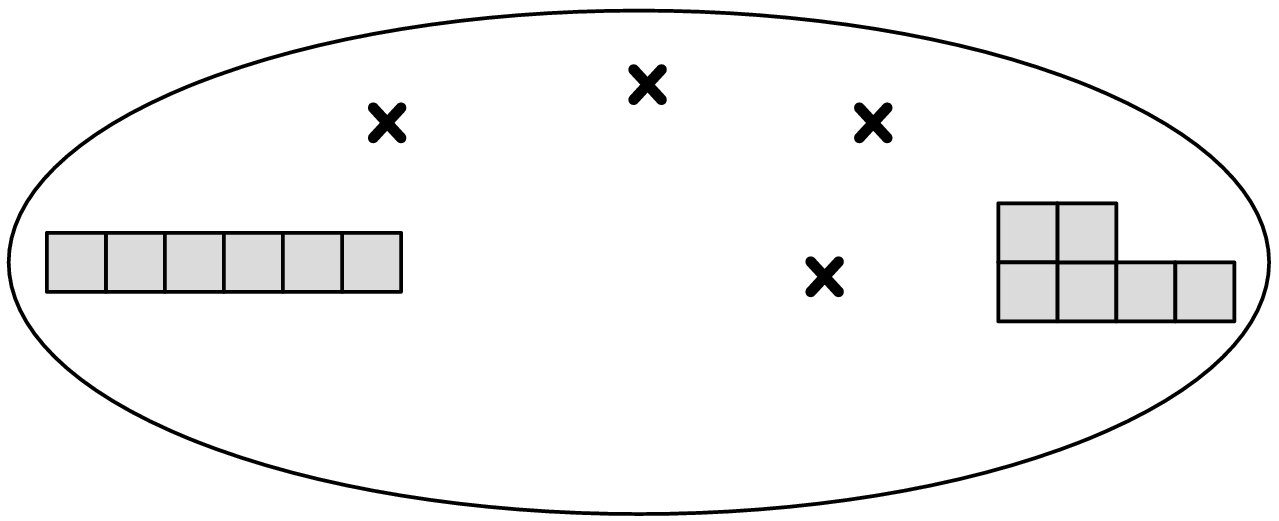}  \\[2em]
\inc{\sizeB}{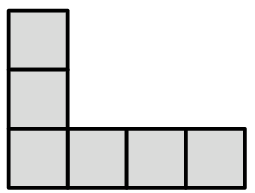} &&  \SO(3) & \inc{\sizeA}{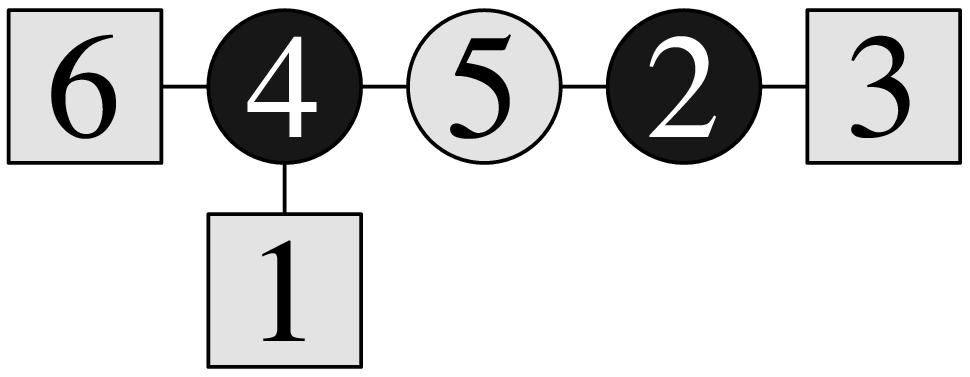} & \inc{\sizeA}{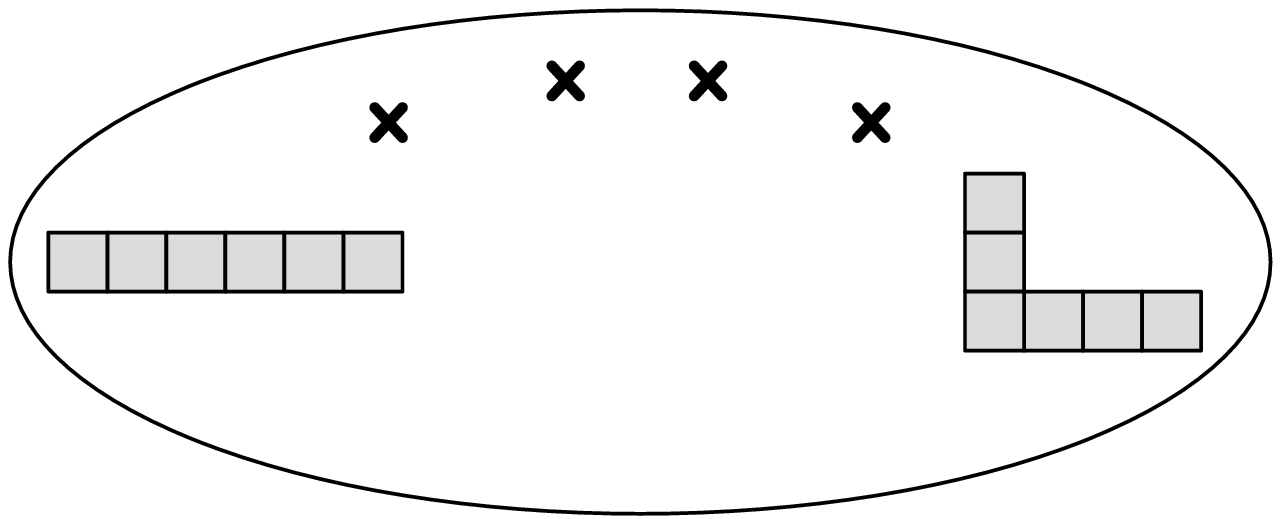} \\[2em]
\inc{\sizeB}{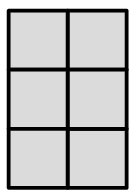} &&\SO(2) & \inc{\sizeA}{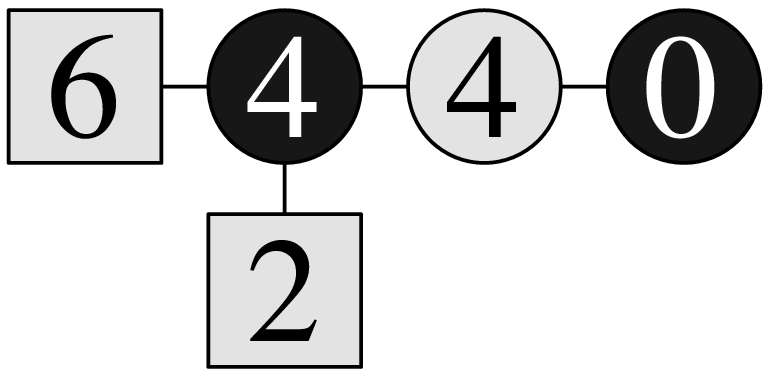} & \inc{\sizeA}{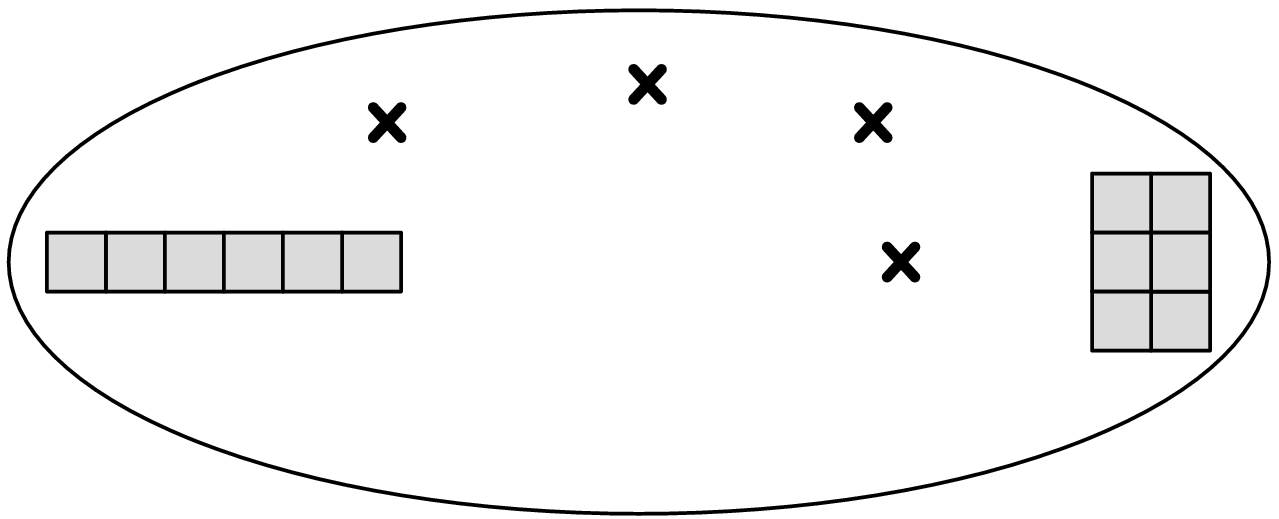} \\[2em]
\inc{\sizeB}{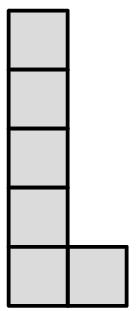} && \text{none} &\inc{\sizeA}{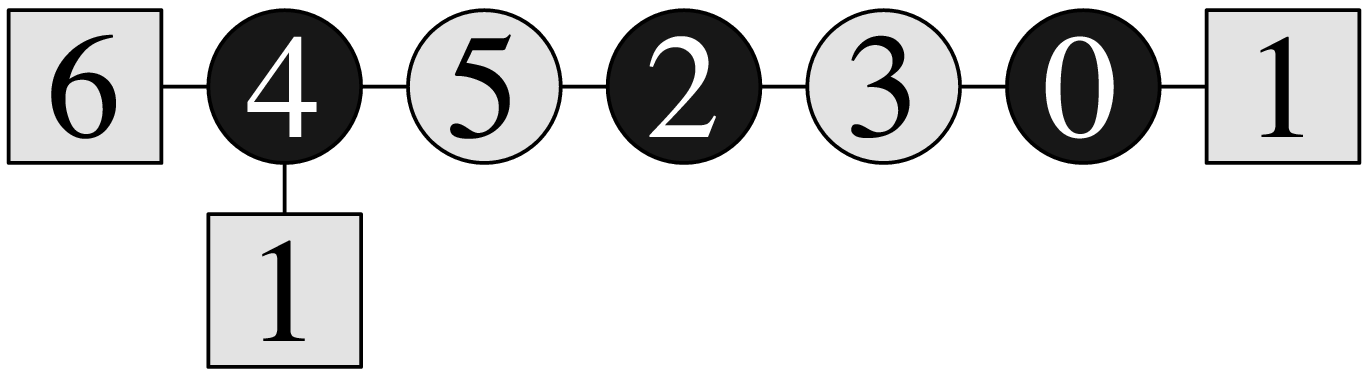} & \inc{\sizeA}{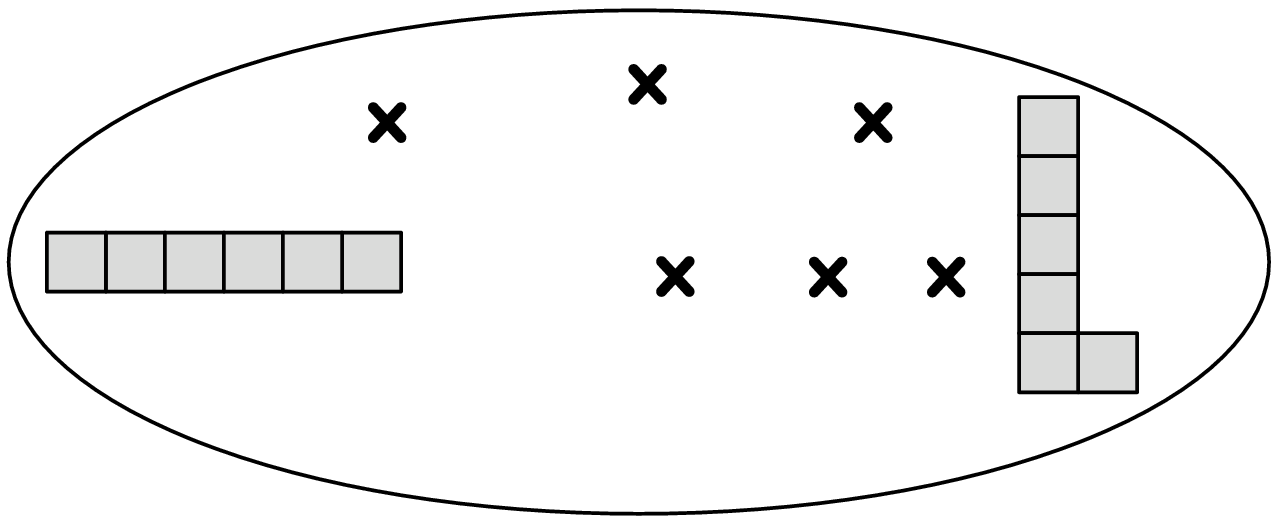}
\end{array}
\]
\caption{One class of punctures of the six-dimensional $D_N$ theories are marked by
$\SO$ tableaux, which encode   embeddings of $\SU(2)$ into $\SO(2N)$.
On the right of the each tableau are the associated flavor symmetry and
a quiver theory whose G-curve has a corresponding puncture.
A puncture whose tableau consists of one row of width $2N$ is
 a full puncture $\odot$.
Quivers with $\USp(0)$ `gauge group' need to be understood as a shorthand
for the brane configurations,  as explained in the text. \label{SOtails}}
\end{table}

\begin{table}
\[
\begin{array}{@{\extracolsep{1ex}}c@{\hskip1ex}|c@{\hskip1ex}|c@{\hskip1ex}||c@{\hskip1ex}|c}
\text{Tableau} & \text{Alias} & \text{Flavor} & \text{Quiver} & \text{G-curve} \\
\hline
\inc{\sizeB}{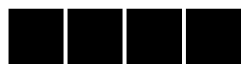} & \star &\USp(4)  & \inc{\sizeA}{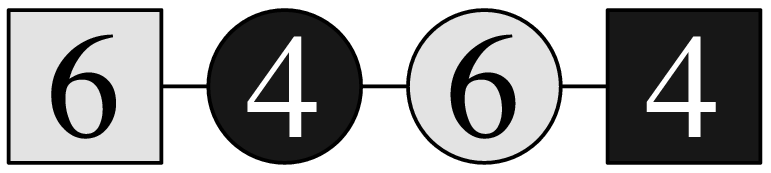} & \hbox{\vbox to 2em{}}\inc{\sizeA}{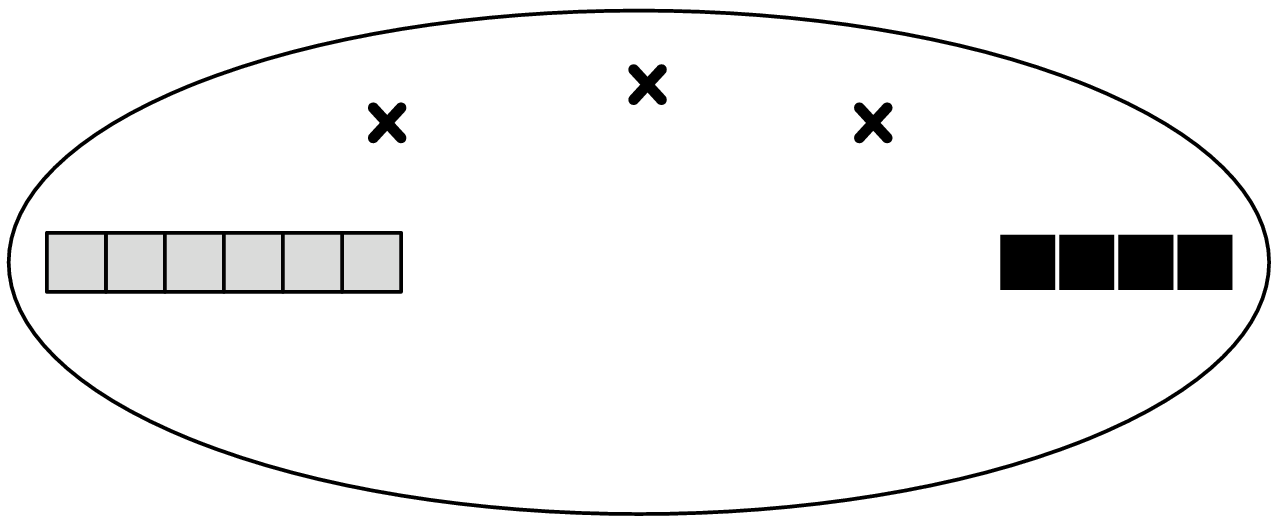} \\[2em]
 \inc{\sizeB}{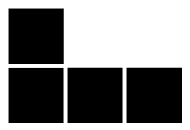} & &\USp(2) & \inc{\sizeA}{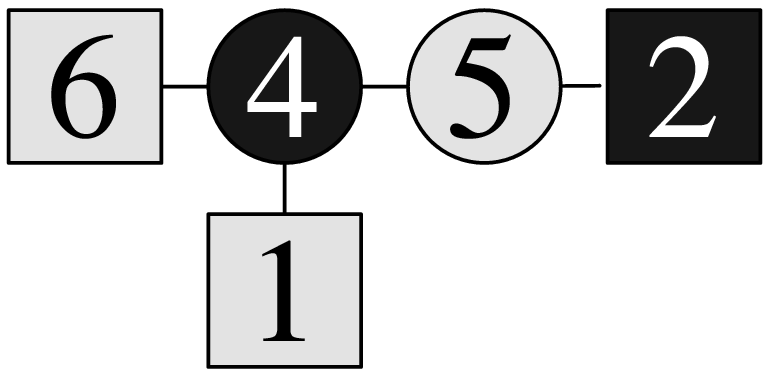} & \inc{\sizeA}{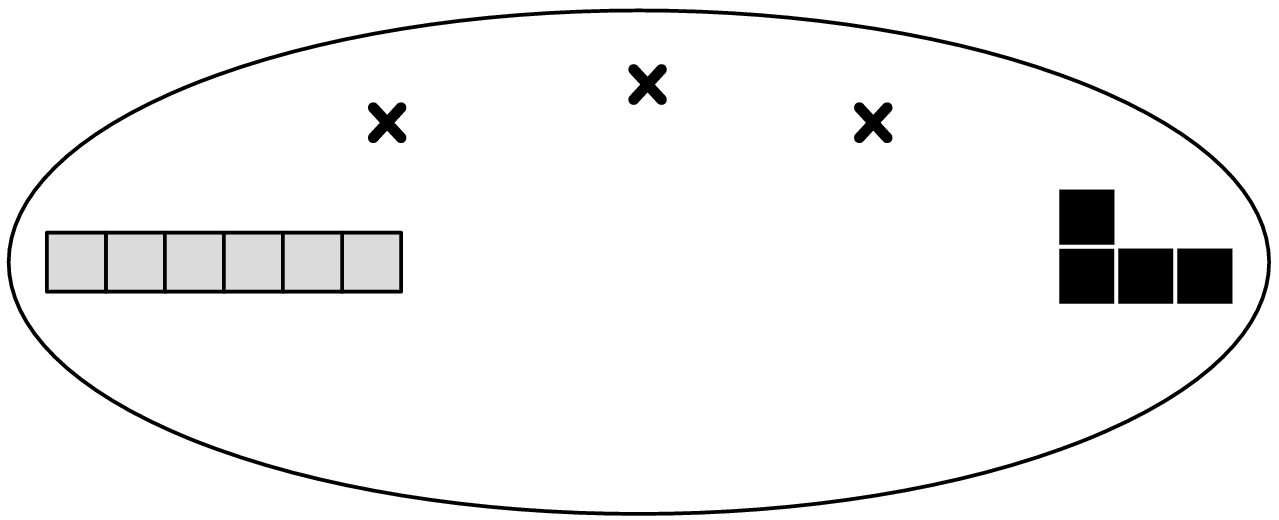} \\[2em]
\inc{\sizeB}{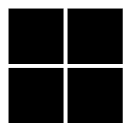} & &\SO(2) & \inc{\sizeA}{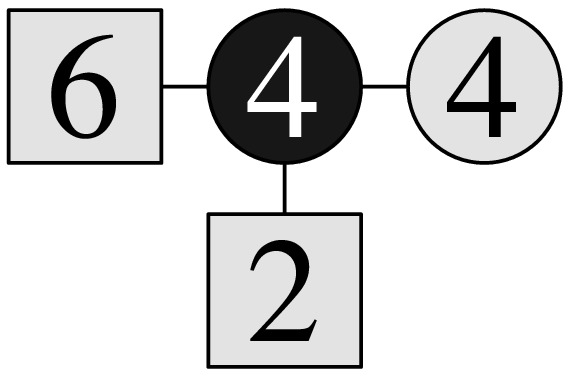} &  \inc{\sizeA}{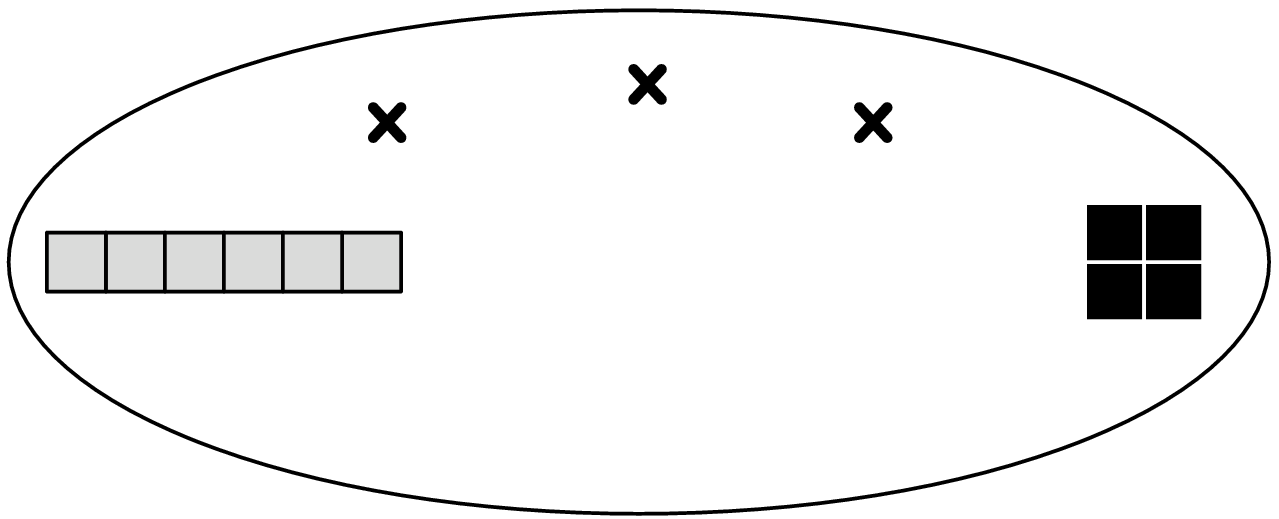}  \\[2em]
\inc{\sizeB}{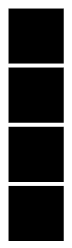} &  \times &\text{none} & \inc{\sizeA}{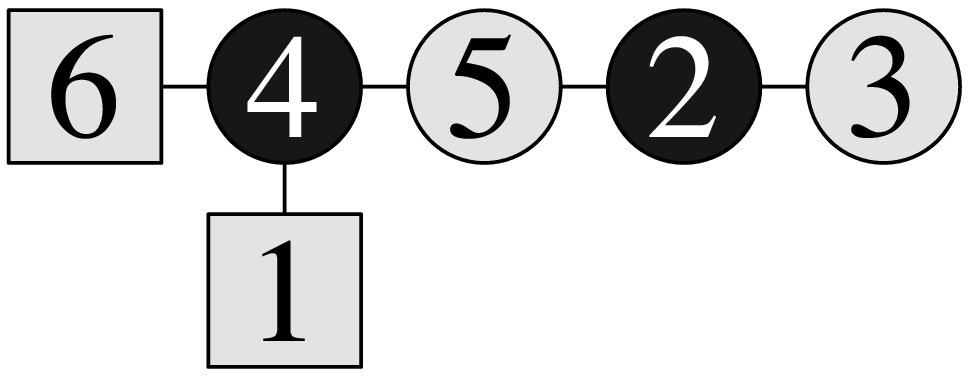} & \inc{\sizeA}{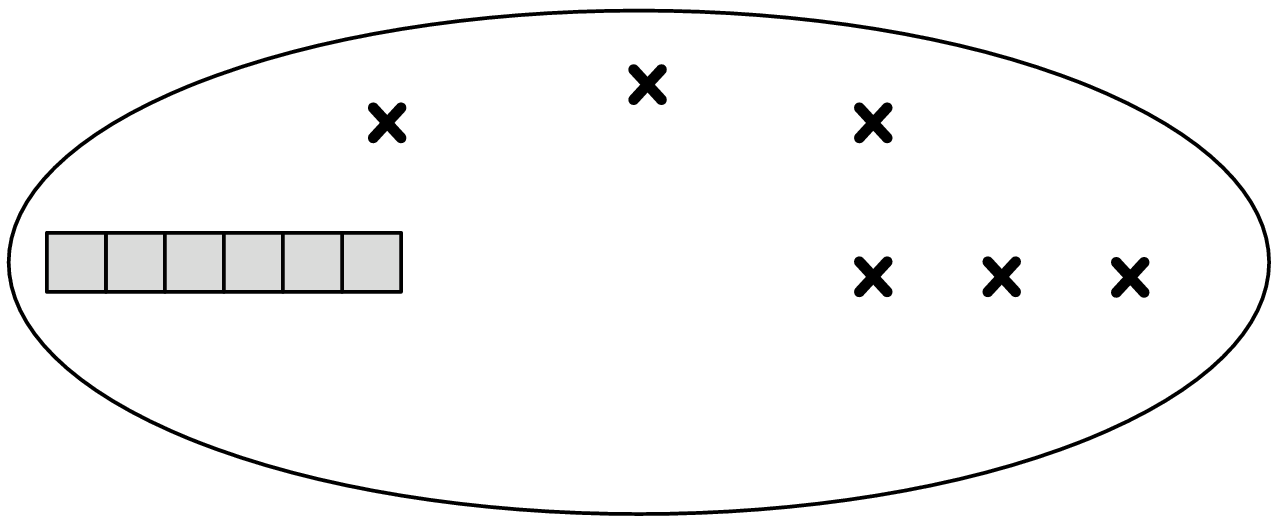} 
\end{array}
\]
\caption{
Another class of punctures of the six-dimensional $D_N$ theories are marked by
$\USp$ tableaux, which encode $\SU(2)$ embeddings into $\USp(2N-2)$.
On the right of the each tableau are the associated flavor symmetry and
a quiver theory whose G-curve has a corresponding puncture.
A puncture whose tableau consists of one column of height $2N-2$ is a simple puncture $\times$,
and a puncture whose tableau consists of one row of width $2N-2$ is a full puncture $\star$.
\label{USptails}}
\end{table}

Let $l_h$ be the number of columns of height $h$ in a given tableau. 
One finds in general that \begin{itemize}
\item  $l_h$ for even $h$ is even for a gray tableau;
it just guarantees that  $d_a$  be even for $\USp$ gauge groups.
Then one can associate 
an  embedding of $\SU(2)$ into $\SO(2N)$, $\rho:\SU(2)\to\SO(2N)$,
given by the decomposition
of the vector representation $2N$ of $\SO(2N)$  under $\SU(2)$ via \begin{equation}
2N \to \underbrace{1+1+\cdots+1}_{l_1} + \underbrace{2+\cdots+2}_{l_2} + \cdots .
\end{equation} 
Recall that the irreducible representation of $\SU(2)$ of dimension $h$
for even $h$ is pseudo-real.
The embedding above is possible
because  $l_h$ copies of this irreducible representation 
can be strictly real, $l_h$ being even.
Thus we call such a tableau an $\SO(2N)$ tableau.
\item  Similarly, $l_h$ for odd $h$ is even for a black tableau. 
Again, this just guarantees that  $d_a$  is even for $\USp$ gauge groups.
Let us then associate an  embedding $\rho$ of $\SU(2)$
into $\USp(2N-2)$, given by the decomposition
of the fundamental representation $2N-2$ of $\USp(2N-2)$ under $\SU(2)$ via \begin{equation}
2N-2 \to \underbrace{1+1+\cdots+1}_{l_1} + \underbrace{2+\cdots+2}_{l_2} + \cdots .
\end{equation} 
Recall that the irreducible representation of $\SU(2)$ of dimension $h$
for odd $h$ is strictly real.
The embedding above is possible
because  $l_h$ copies of this irreducible representation 
can be pseudo-real, $l_h$ being even.
Thus we call such a tableau a $\USp(2N-2)$ tableau.
\end{itemize}

In this way, we associate a tableau for each superconformal tail,
which encodes its information concisely.
One can observe the following facts concerning superconformal tails
and the flavor symmetries associated to them:
\begin{itemize}
\item 
For a tail ending in a $\USp$ group, 
the flavor symmetry associated to it is  \begin{equation}
\prod_{h:\text{odd}, l_h\ge 2} \SO(l_h) \times
\prod_{h:\text{even}, l_h\ge 2} \USp(l_h), \label{SOcommutant}
\end{equation} which coincides  exactly with  the commutant inside $\SO(2N)$ 
of the $\SU(2)$ embedding associated to the tableau labeling the tail.
\item  Similarly, for a tail ending in an $\SO$ group,
the flavor symmetry is \begin{equation}
\prod_{h:\text{odd}, l_h\ge 2} \USp(l_h) \times
\prod_{h:\text{even}, l_h\ge 2} \SO(l_h),  \label{USpcommutant}
\end{equation}which agrees with  the commutant  inside $\USp(2N-2)$ of
the $\SU(2)$ embedding associated to the tableau.
\end{itemize}

\begin{figure}
\[
\inc{\sizeB}{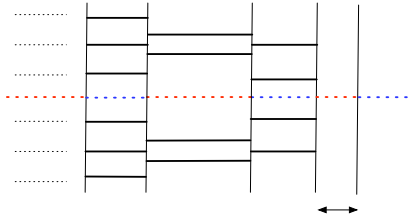}
\]
\caption{Brane configuration involving ``$\USp(0)$'' part.
It would correspond to a quiver with gauge groups
$\cdots\times\SO(6)\times\USp(4)\times\SO(4) \times\text{``$\USp(0)$''}$.\label{usp0}}
\end{figure}

Conversely, given a $\USp(2N-2)$ tableau one can always construct a superconformal tail
ending in an $\SO$ gauge group, and given an $\SO(2N)$ tableau one can write down
a tail ending in a $\USp$ gauge group.
However, there is one class of exceptions
which are $\SO(2N)$ tableaux consisting of just two columns,
associated to the decomposition \begin{equation}
2N \to (2N-k) + k.
\end{equation} Here $k$ is  odd unless  $N$ is even, in which case $k=N$ is also allowed.
Naive application of the algorithm above associates a superconformal tail of the form \begin{equation}
\cdots \times \SO(6) \times \USp(4)\times \SO(4) \times \text{``$\USp(0)$''}
\end{equation} which does not make sense in a purely gauge-theoretic setting.
However, as was the case for $\SU$ quivers, one can still write down a brane configuration
corresponding to this situation (Fig~\ref{usp0})
and consistently lift it to M-theory.\footnote{%
This situation might be related to the appearance for $\USp(0)$ gauge group
of the gluino condensate in the framework of Dijkgraaf-Vafa \cite{Intriligator:2003xs},
which was later interpreted as the stringy D-brane instanton contributions  
in \cite{GarciaEtxebarria:2008iw}.
It would be interesting to clarify the relation; the author thanks Masaki Shigemori and I\~naki Garc\'\i a-Etxebarria for discussions.}

Finally let us discuss the behavior of  $\varphi_{2k}$ and $\varphi_{\tilde N}$ 
at the punctures, which can be found by a careful analysis
of the Seiberg-Witten curves.
One  finds that it is not sufficient to specify 
the degrees of the poles for each $\varphi_{2k}$ or $\varphi_{\tilde N}$,
contrary to the case of the $A_N$ theory.
For example, at the puncture associated to the $\SO(6)$ tableau
corresponding to $6\to 3+3$, we find the following two conditions,
whose derivation can be found in Appendix~\ref{curves}:
\begin{itemize}
\item $\varphi_2$, $\varphi_{\tilde 3}$ and $\varphi_{4}$ have poles
of degree $1,1,2$ respectively; and 
\item $(\varphi_2)^2-4\varphi_4$  has only a simple pole.
\end{itemize}
The second condition guarantees that the M5-brane wrapping the Seiberg-Witten curve
does not have a single, transversal intersection with the M-theory orientifold.

This sounds slightly puzzling, considering the fact that 
the $D_3$ theory in six dimensions is equivalent to the $A_3$ theory,
for which the defects were classified and such a polynomial constraint was not found.
Indeed, the decomposition $6\to 3+3$ of the $\mathbf{6}$ of $\SO(6)$
corresponds to $4\to 3+1$ of the  $\mathbf{4}$ of $\SU(4)$, for which the pole structure was just that
all $\phi_{2,3,4}$ have simple poles at the puncture, see Table~\ref{SU4quivers}.
But upon  further reflection this is exactly what is expected. Say that an element
of the Cartan subalgebra of $\SO(6)$ acts on the $\mathbf{6}$ as
\begin{equation}
\mathrm{diag}(a,-a,b,-b,c,-c);
\end{equation} then it acts on the $\mathbf{4}$ as \begin{equation}
\mathrm{diag}(a+b+c,a-b-c,-a+b-c,-a-b+c)/2.
\end{equation} Therefore $\phi_4$ and $\varphi_4$ cannot be just equated; instead they satisfy
\begin{equation}
2\phi_2=\varphi_2, \qquad
16\phi_4= \varphi_2^2 - 4\varphi_4.
\end{equation}
Thus we find that the condition on the worldvolume fields as found from the $D_3$ theory
is the same as the one found from the point of view of the $A_3$ theory.

\subsection{Duality with SCFT with $E_7$ flavor symmetry}
\begin{figure}
\[
\begin{array}{cc@{\qquad}c}
\text{before:} &  \inc{\sizeA}{q3111g} & \inc{\sizeA}{r3111g} \\[2em]
&\Downarrow  & \Downarrow \\
\text{after:} & \inc{\sizeA}{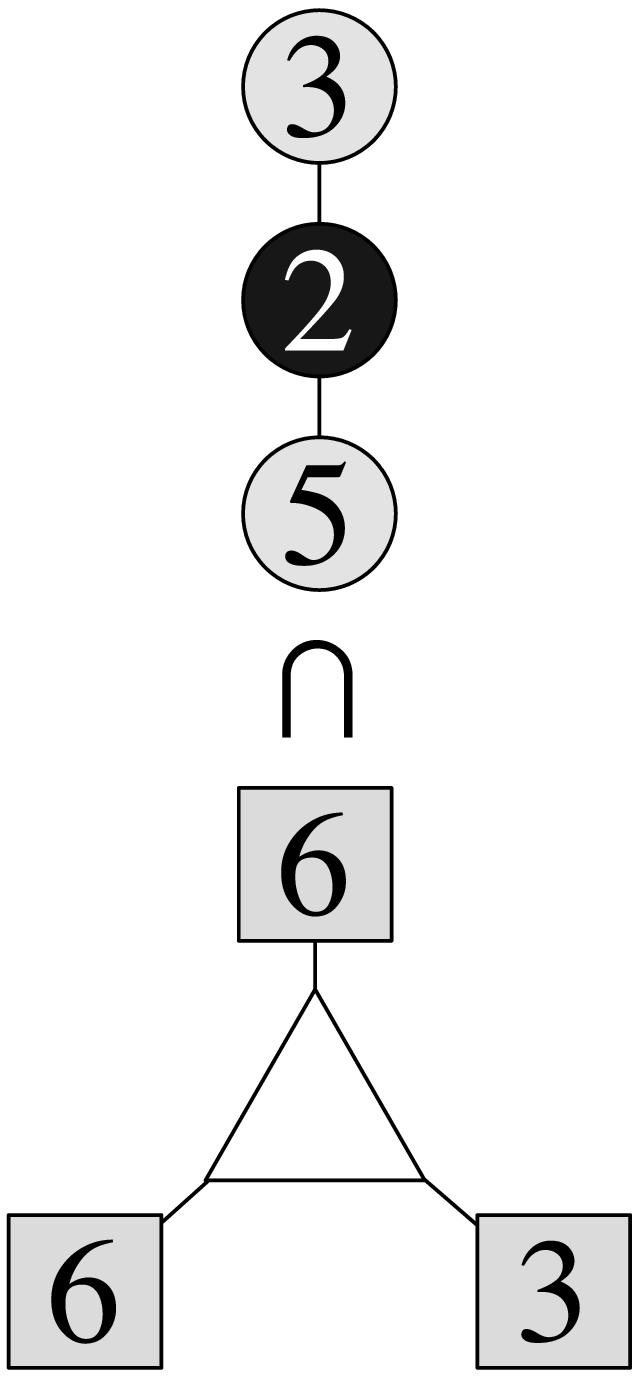} & \inc{\sizeA}{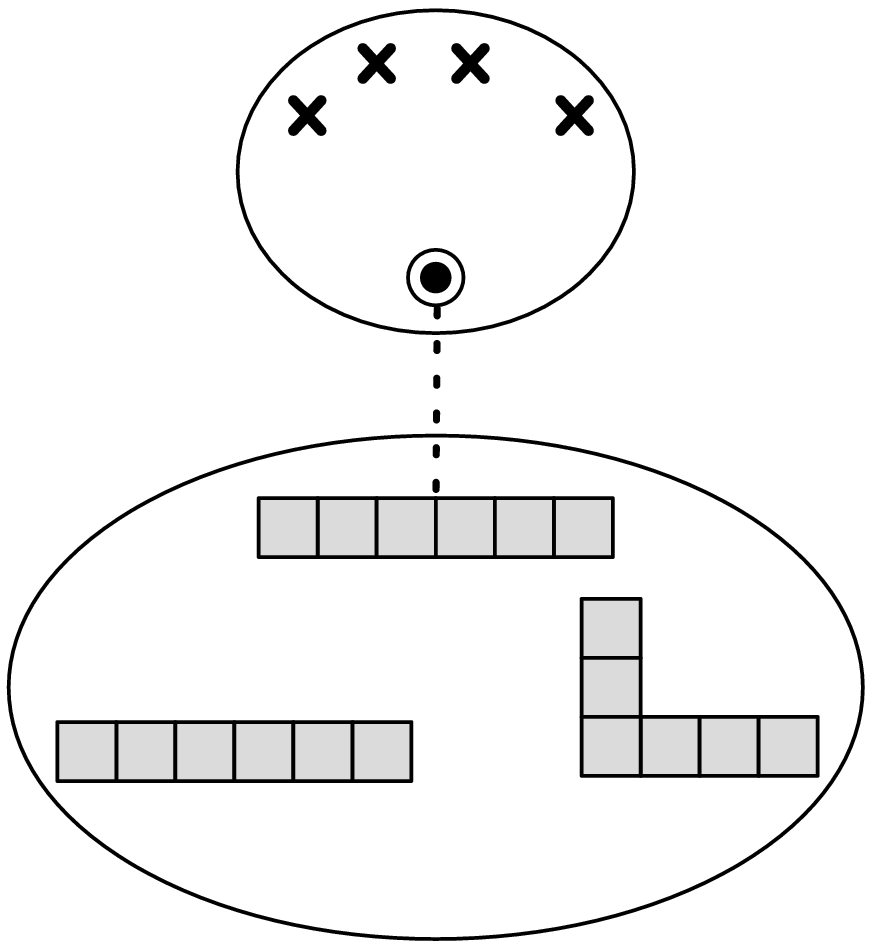} 
\end{array}
\]
\caption{S-duality involving the SCFT 
with $E_7$ flavor symmetry\label{ASE7}}
\end{figure}
Having analyzed general punctures of the $D_N$ theory,
we can now have some more fun.
Let us start from the quiver with gauge groups $\USp(4)\times \SO(5)\times \USp(2)$,
shown in the first row of Fig.~\ref{ASE7}.
As before, we can go to a region where four punctures of type $\times$
collide, decoupling a tail with gauge groups $\SO(5)\times \USp(2)\times \SO(3)$.
The resulting strongly-coupled theory
has no marginal coupling because there are only three punctures on the sphere.
There is only one Coulomb branch operator and its dimension is four,
because the original theory contained three operators of dimension two
and two of dimension four,
whereas  the decoupled tail has three dimension-two
and one dimension-four operators.

\begin{figure}
\[
\inc{\sizeA}{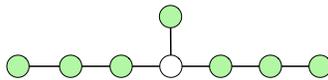}
\]
\caption{Extended Dynkin diagram of $E_7$
showing subgroup $\SU(4)^2\times \SU(2)$ \label{e7dynkin}}
\end{figure}

This suggests that this theory is the $E_7$ SCFT 
of Minahan-Nemeschansky \cite{Minahan:1996cj}.
One can perform many tests of the proposal: one can easily check that
the central charges $a$ and $c$ agree with what were found in \cite{Argyres:2007cn}; 
and the flavor symmetry manifest  in this description is naturally a subgroup of $E_7$,
\begin{equation}
\SO(6)^2\times \SO(3)\simeq \SU(4)^2\times \SU(2) \subset E_7.
\end{equation} 
This subgroup comes from the decomposition of the extended Dynkin diagram,
see Fig.~\ref{e7dynkin}. The flavor symmetry central charges of this subgroup
agree with those of $E_7$, which were found in \cite{Argyres:2007cn}.

The S-duality found in \cite{Argyres:2007cn} involving the $E_7$ SCFT
started from $\USp(4)$ gauge theory with twelve half-hypermultiplets, which is
exactly the  first example in Table~\ref{SOtails}.
The infinitely strongly-coupled limit
corresponds to collapsing two singularities of type $\times$.
The analysis above indicates that this procedure results in one Young tableau
with columns of height $3,1,1,1$. It would be fruitful to analyze 
which defects can arise when two defects  of general type collide.
Such collisions should provide a wealth of new S-dualities.

\section{Discussion}\label{discussion}

In this paper, we generalized the construction of \cite{Gaiotto:2009we}, 
which realized many four-dimensional SCFTs as compactifications
of the six-dimensional $A_{N-1}$ theory,
to the $D_N$ theory. 
We utilized this construction to find a new class of isolated SCFTs with $\SO(2N)^3$ flavor symmetry,
which arise in  strongly-coupled limits of linear quivers of $\SO$ and $\USp$ gauge groups. We also saw how  the $E_7$ SCFT of Minahan and Nemeschansky arises 
from this construction.

In \cite{Gaiotto:2009we}, it was noted that the types of tails
of $\SU$ superconformal quivers for six-dimensional $A_{N-1}$ theory
can be naturally associated to Young tableaux;
we can naturally associate an  embedding of $\SU(2)$ 
into $\SU(N)$ to such Young tableau,
whose commutant inside $\SU(N)$ gave the flavor symmetry of that tail.
In this paper, the analysis was extended to alternating $\SO$--$\USp$ quivers
for the six-dimensional $D_N$ theory
and it was found that the tails of such quivers can naturally be associated to 
an  embedding of $\SU(2)$ into
either  $\SO(2N)$ or  $\USp(2N-2)$;
again, the flavor symmetry associated to the tail is given by the commutant of 
that embedding of $\SU(2)$.

We also saw that the simplest  kinds of defects of the $D_N$ theory
have  $\bZ_2$ monodromy for the Pfaffian operator. 
This is suggestive in that the Pfaffian is odd under
the outer automorphism of the $D_N$ Lie algebra,
whose quotient is exactly $\USp(2N-2)$,
which was used in the labeling of the tails. It would be interesting to consider
defects associated to other outer automorphisms of $A_{N-1}$ or $D_4$,
and identify their realizations using quiver gauge theory.

The most pressing issue is to find out how the association to the defects of 
an  embedding of $\SU(2)$
into $\SU$, $\SO$ or $\USp$ groups
can be intrinsically understood from a six-dimensional point of view,
and how these embeddings control the behavior of the scalar fields around them.
These defects are of codimension two. Therefore, if we
compactify the six-dimensional $(2,0)$ theory on a torus
parallel to the worldvolume of the defects,
we obtain surface operators of the $\cN=4$ super Yang-Mills in four dimensions.
The study of such surface operators was initiated in\cite{Gukov:2006jk}.
There, it was found that embeddings of $\SU(2)$  naturally appear
which specify the singular part of the field configuration around the defect.
Therefore, the problem seems to be in identifying which of the possible 
defects of four-dimensional $\cN=4$
super Yang-Mills descend from those of six-dimensional theories.
It would be interesting to pursue this relation further.

\section*{Acknowledgments}
It is a great privilege for the author 
to thank Davide Gaiotto for many illuminating discussions.
The author is also indebted to Alfred D. Shapere for careful reading of 
and many detailed comments on the manuscript.
He also benefited from discussions with L. Fernando Alday, I\~naki Garc\'\i a-Etxebarria, Juan M. Maldacena, Nathan Seiberg, Masaki Shigemori and Yoske Sumitomo.
The author is supported in part by the NSF grant PHY-0503584, and in part by the Marvin L. 
Goldberger membership at the Institute for Advanced Study.

\appendix

\section{$\SO(4)$--$\USp(2)$ quivers}\label{so4}
\begin{figure}
\[
\begin{array}{c@{\qquad}c@{\qquad}c}
1) &\inc{\sizeA}{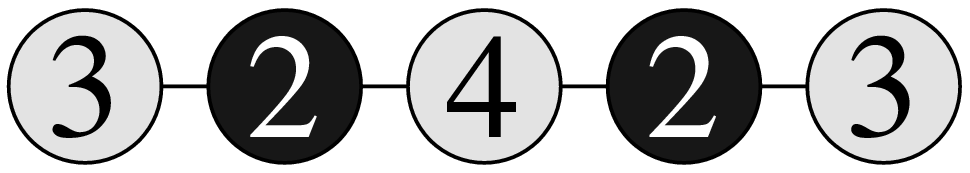} & \inc{\sizeA}{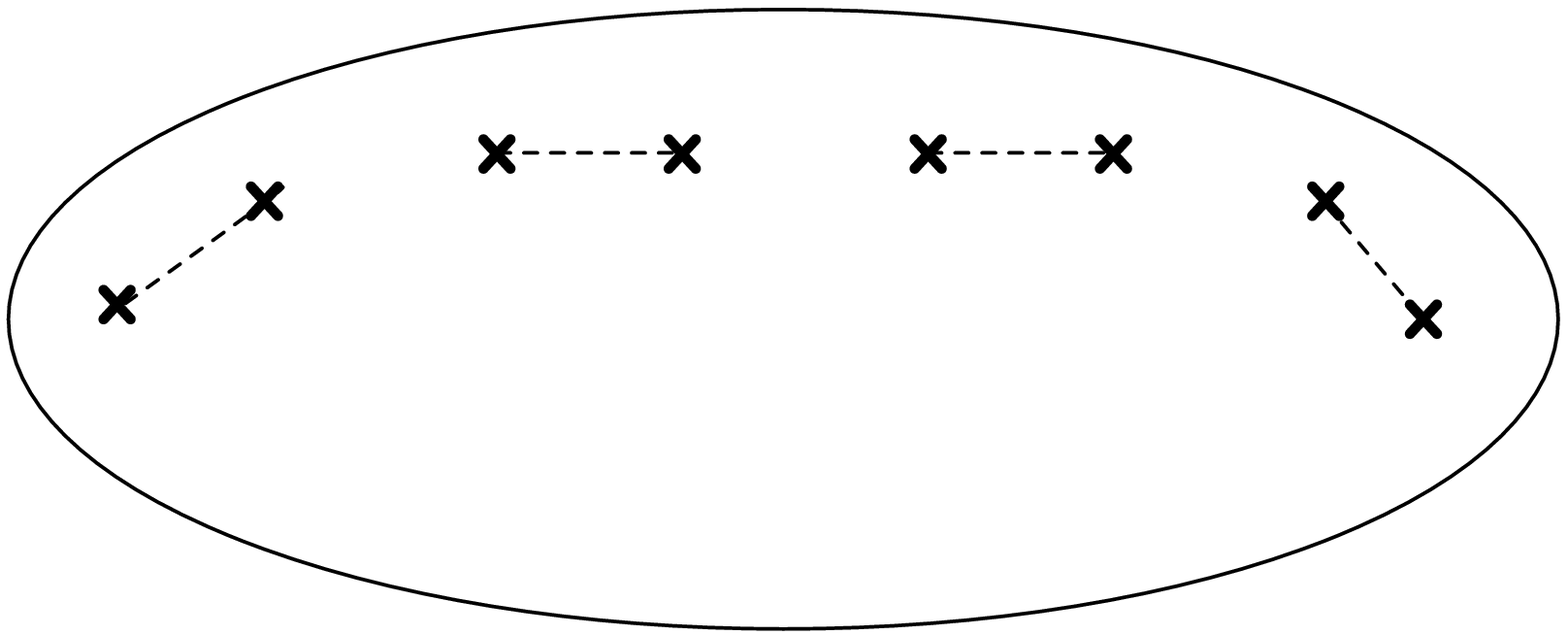}\\[3em]
2) & \inc{\sizeA}{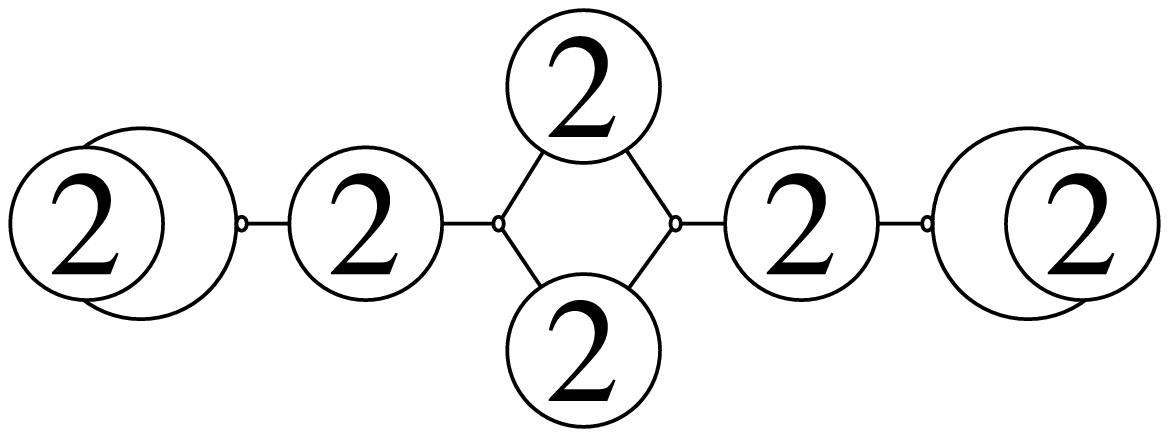} & \inc{\sizeA}{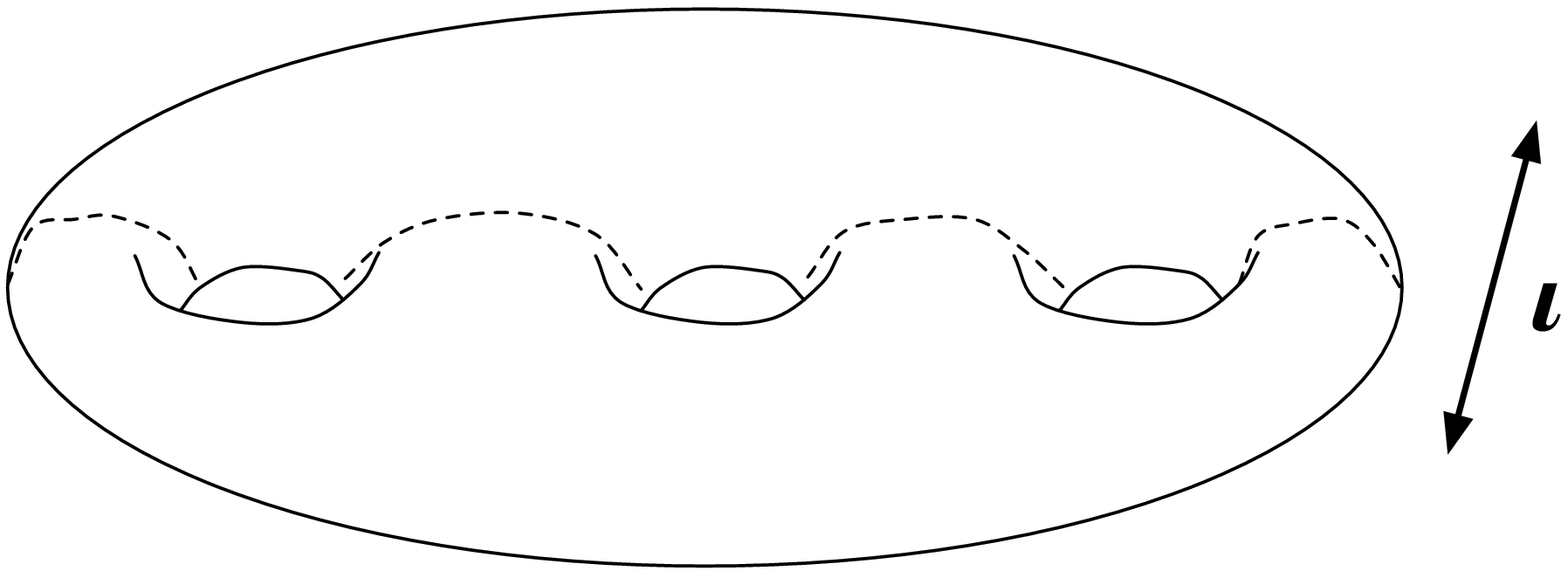}
\end{array}
\]
\caption{An $\SO(4)$--$\USp(2)$ quiver and the same quiver as an $\SU(2)$ generalized quiver.
Corresponding G-curves are also shown. In the first description the curve is a sphere with eight punctures;
in the second it is a hyperelliptic genus-three curve.
\label{so4quiver}}
\end{figure}

It is instructive to analyze the simplest case of the $\SO$--$\USp$ quiver,
namely the case with $\SO(4)$--$\USp(2)$  using our formalism.
Consider a linear quiver theory with the gauge group \begin{equation}
\SO(3)\times\USp(2)\times \SO(4)\times \cdots \USp(2)\times \SO(3),
\label{so4so4}
\end{equation} with $s$ $\SO(4)$, $s+1$ $\USp(2)$ and two $\SO(3)$ factors.
The case $s=1$ is shown in the diagram 1) of Fig.~\ref{so4quiver}.
Following the procedure explained in the main part,
we find that the Seiberg-Witten curve is specified by the G-curve $\Sigma=\CP^1$ 
with $2s+6$  punctures of the same type;
we have two quadratic differentials on $\Sigma$, 
$\varphi_2$ and $\varphi_{\tilde 2}$ corresponding to two Casimirs of $\SO(4)$,
namely the trace of the square and the Pfaffian.
At each puncture, $\varphi_2 \sim (dt)^2/t$ and $\varphi_{\tilde 2} \sim (dt)^2/t^{1/2}$ 
where $t$ is a local coordinate for which the puncture is at $t=0$.

Now let us recall that $\SO(4)\simeq \SU(2)\times \SU(2)$, and the vector representation
of $\SO(4)$ is the tensor product of the doublets of each of the $\SU(2)$ factors;
also that $\SO(3)\simeq \SU(2)$  and the vector representation of $\SO(3)$ is in the 
tensor product of two doublets. We neglect possible issues coming from
the global structure of the groups. This should not cause any problems
as long as we consider theories on the flat $\bR^4$.

Then the quiver can also be  presented as in the diagram 2) of Fig.~\ref{so4quiver}
in the notation of \cite{Gaiotto:2009we}. In this case the G-curve
is  a genus-$(s+2)$ Riemann surface $\Xi$ and there is a quadratic differential 
$\phi_2$ on it. An important constraint is that in the description as an $\SO(4)$--$\USp(2)$ quiver,
we cannot independently vary the coupling constants of two $\SU(2)$ factors of $\SO(4)$.
It is natural to guess that this requirement
translates to the fact that the curve $\Xi$ is hyperelliptic.
Indeed, it gives the correct number of marginal coupling constants
because the number of the moduli of hyperelliptic curves of genus $s+2$
is $2s+3$, which agrees with the number of gauge factors
in the quiver \eqref{so4so4}.

Now $\Xi$ is equipped with the hyperelliptic involution $\iota$ which flips the two sheets;
the fixed points are exactly the branch points on $\Sigma$. We can split $\phi_2$ on $\Xi$
into even and odd parts under $\iota$: \begin{equation}
\phi_2= \phi_2^+ + \phi_2^-.
\end{equation} 
We then regard $\phi_2^\pm$ as differentials on $\Sigma$.

Pick a puncture on $\Sigma$ and take the local coordinate $t$ so that the puncture is at $t=0$.
The local coordinates on $\Xi$ is then $s=t^{1/2}$.
$\phi_2^+$ is holomorphic and even in $s$,
which translates to the condition
\begin{equation}
\phi_2^+ \sim (ds^2) \sim (dt^2)/t,
\end{equation} implying that $\phi_2^+$ has a simple pole at the branch points.
Similarly, $\phi_2^-$ behaves as \begin{equation}
\phi_2^- \sim s(ds^2) \sim (dt^2)/t^{1/2}.
\end{equation}
Therefore we can identify $\varphi_2$ with $\phi_2^+$
and $\varphi_{\tilde 2}$ with $\phi_2^-$.

\section{Curves for general $\SO$--$\USp$ quivers}\label{curves}

Here we provide some details of the derivation of the Seiberg-Witten curves
for general linear quiver gauge theories 
with alternating $\SO$ and $\USp$ gauge groups.
The brane construction was reviewed in Sec.~\ref{brane},
see Fig.~\ref{braneconfig} for a drawing of the system.

Let us first recall how D6 branes lift to a Taub-NUT space in M-theory.
Let $N_f$ be the total number of D6 branes.
When all of them are at $x^4=x^5=0$,
the resulting Taub-NUT space is given as a complex manifold by the equation
\begin{equation}
yz = v^{N_f},
\end{equation}  and the orientifolding in M-theory acts by sending $v\to -v$.
The action of  orientifolding  on $y$ and $z$ depends on the quiver;
for simplicity we assume that $y$ is fixed for now.

\begin{figure}
\[
\inc{\sizeC}{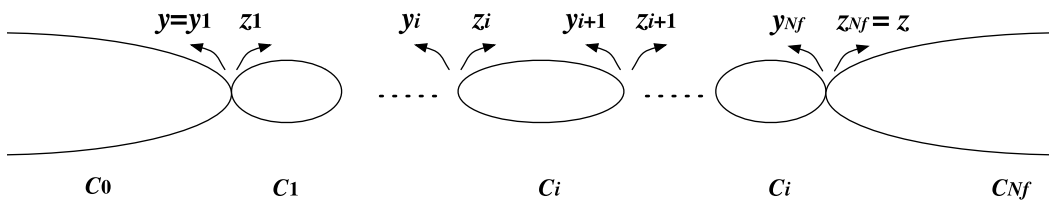}
\]
\caption{Schematic description of the Taub-NUT space, the lift of D6-branes.
$C_{1}$ to $C_{N_f}$ are blown-up two-cycles. \label{taub}}
\end{figure}

The origin $y=z=v=0$ is blown up as long as the positions
of D6-branes along $x^6$ directions are distinct.
The blown-up, smooth manifold is given by introducing extra pairs of local coordinates
$(y_i,z_i)$ at the $i$-th nut $i=1,2,\ldots, N_f$,  such that
\begin{equation}
y_1z_1=y_2z_2= \cdots = v, \label{foo}
\end{equation} and \begin{equation}
y=y_1,\quad z_1=1/y_2, \quad z_2=1/y_3, \quad \ldots, \quad z_{N_f}=z.\label{bar}
\end{equation}
There are $N_f-1$ $\CP^1$'s parametrized by $y_{2,3,\ldots,N_f}$
which we call $C_i$, see Fig.~\ref{taub}. We analogously define
the loci $z=0$ and $y=0$ to be $C_0$, $C_{N_f}$.
The relations \eqref{foo}, \eqref{bar} imply that the orientifolding
fixes $C_\text{even}$ but that it acts  by multiplication by $-1$ on $C_\text{odd}$.
Therefore the M-theory orientifold exists at $C_\text{even}$, but not at $C_\text{odd}$.

Now let us consider an $\SO$--$\USp$ superconformal quiver gauge theory 
for the $D_N$ theory, specified by integers $(d_a)$, $(\delta_a)$
and $(k_a)$, see Sec.~\ref{brane} for notations.
The Seiberg-Witten curve is  a curve in
the Taub-NUT space discussed above \cite{Landsteiner:1997vd,Brandhuber:1997cc},
given by \begin{multline}
F(v,t)=v^{2N} t^{n+1} + v^{2N-d_1} P_1(v) t^n
+ v^{2N-d_2} P_2(v) t^{n-2} + \cdots \\
+ v^{2N-d_n} P_n(v) t + v^{2N}=0.\label{generalSWcurve}
\end{multline} 
where $t=y_k$ and $k$ is the number of D6-branes of  the left hand tail
of the superconformal quiver.
$P_a(v)$ is a polynomial of degree $d_a$,
even or odd in $v$ according to the parity of $d_a$:
\begin{equation}
P_a(v)=c_a v^{d_a}  + u_a^{(2)} v^{d_a-2} + u_a^{(4)} v^{d_a-4} + \cdots.
\end{equation}
In the semi-classical regime, $c_a$  encodes the gauge coupling constant,
whereas $u_a^{(2k)}$ encodes the vacuum expectation values of the adjoint
scalar field of the $a$-th gauge groups, 
{\em except} the constant term of $P_a(v)$
for which the gauge group is $\USp$.
These constant terms are determined by the requirement that the
resulting M5-brane intersects the cycles $C_i$ in a manner
consistent with orientifolding in M-theory.
The main condition is that an M5 brane cannot intersect with the M-theory
orientifold five-plane  transversally; an even number of M5-branes
need to intersect at a point on an orientifold five-plane.
This condition was first formulated in \cite{deBoer:1998by}. Refer to  \cite{Hori:1998iv} for more details.

Let us define $x=v dt/t$ and   \begin{equation}
\varphi_{2k}= 
\frac{u_1^{(2k)}t^{n}+u_2^{(2k)}t^{n-1}+\cdots+ u_n^{(2k)} t }
{\prod (t-t_a)}
\left(\frac{dt}{t}\right)^{2k},\label{bot}
\end{equation}
where we defined $u_a^{(2k)}=0$  when $d_a<2k$,
and $\prod (t-t_a) = t^{n+1} + c_1 t^n + \cdots + 1.$
Then one can rewrite the curve above into Gaiotto's form:
\begin{equation}
0=x^{2N}+ \varphi_{2} x^{2N-2} + \varphi_{4} x^{2N-4} +\cdots + \varphi_{2N}.
\end{equation}

The structure of the poles at the punctures at $t=0,\infty$
can be readily found from the form \eqref{bot},
but the conditions  imposed on the constant terms of $P_a(v)$
by the consistency of the M-theory orientifold
are rather intricate, and the author has not found
a concise way to express them for a general sequence of gauge groups.
Instead they are illustrated through two examples, which were used
in the main part of the paper.

\begin{figure}
\[\begin{array}{c}
\inc{\sizeA}{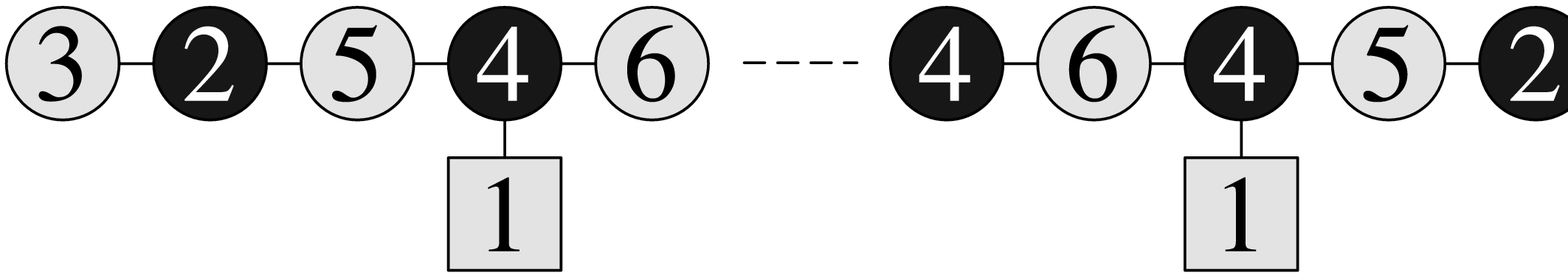}\\[2em]
\inc{\sizeC}{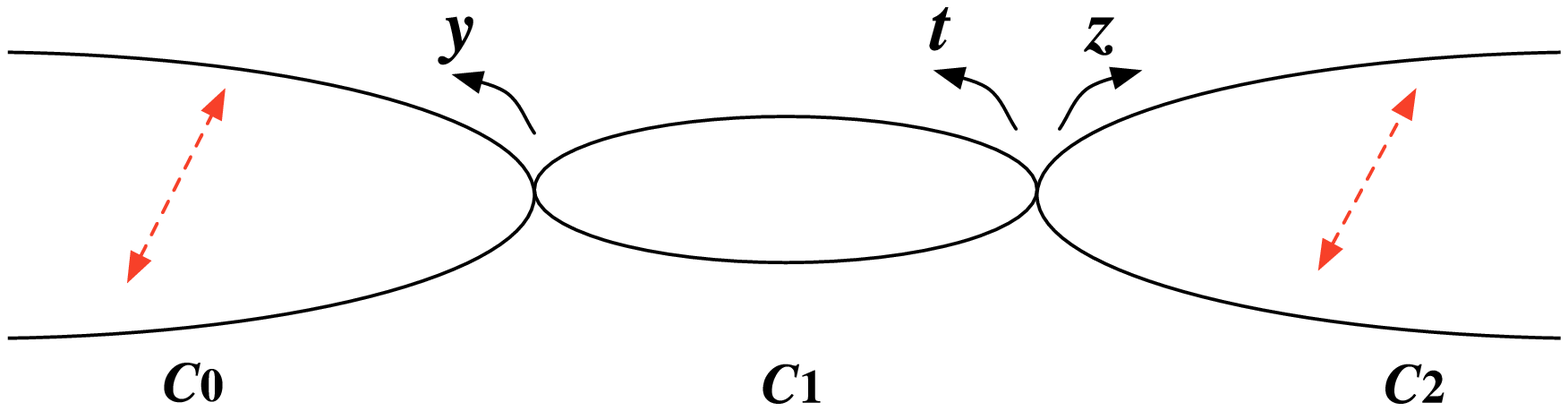}
\end{array}\]
\caption{A quiver, and the corresponding Taub-NUT space in which its Seiberg-Witten curve 
is embedded. Orientifolding fixes $C_1$, but acts as multiplication by $-1$ on
$C_{0}$ and $C_2$. \label{taub-simple}}
\end{figure}

The first example is the quiver drawn in Fig.~\ref{taub-simple},
which was the main topic of Sec.~\ref{SO6pants}.
The Taub-NUT space $yz=v^2$ is given in the right hand side of the same figure.
The red broken arrow on $C_0$ and $C_2$
indicates that the orientifolding sends $y\to -y$ and $z\to -z$.
The Seiberg-Witten curve was given in \eqref{generalSWcurve}.
$t$ is the local coordinate of $C_1$.
Let $n=2b+7$ be the total number of gauge groups; $b$ is the number of 
$\SO(6)$ gauge groups. 
The consistent way to 
eliminate extra parameters in $P_\text{even}$ is then to set 
the constant parts of $P_{2}$, $P_4$, $P_{2b+4}$ and $P_{2b+6}$ to zero,
and to require  \begin{equation}
F(0,t)=\alpha t^5 \prod_{i=1}^{b-1} (t-q_i)^2
\end{equation} for some complex numbers $\alpha$, $q_i$.
Indeed, 
the intersection of $C_1$ with the Seiberg-Witten curve 
is given by the double zeros $q_i$, as required by the consistency of 
the M-theory orientifold \cite{Hori:1998iv}.
Furthermore, 
the intersection with $C_0$ is given by\begin{equation}
u_{1}^{(2)}y^2 - u_3^{(4)} =0
\end{equation} which is compatible with the orientifolding action.
The same can be said for $C_2$.
Under these constraints, one finds that $\varphi_2$, $\varphi_4$, \ldots,
$\varphi_{2N}$ all have simple poles at $t=0$.
This translates to the behavior $\sim 1/t^{1/2}$ for $\varphi_{\tilde N}$.

\begin{figure}
\[
\inc{\sizeA}{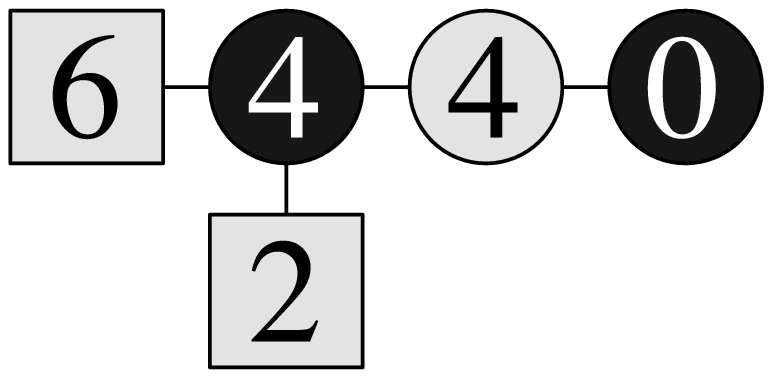}\qquad
\inc{\sizeC}{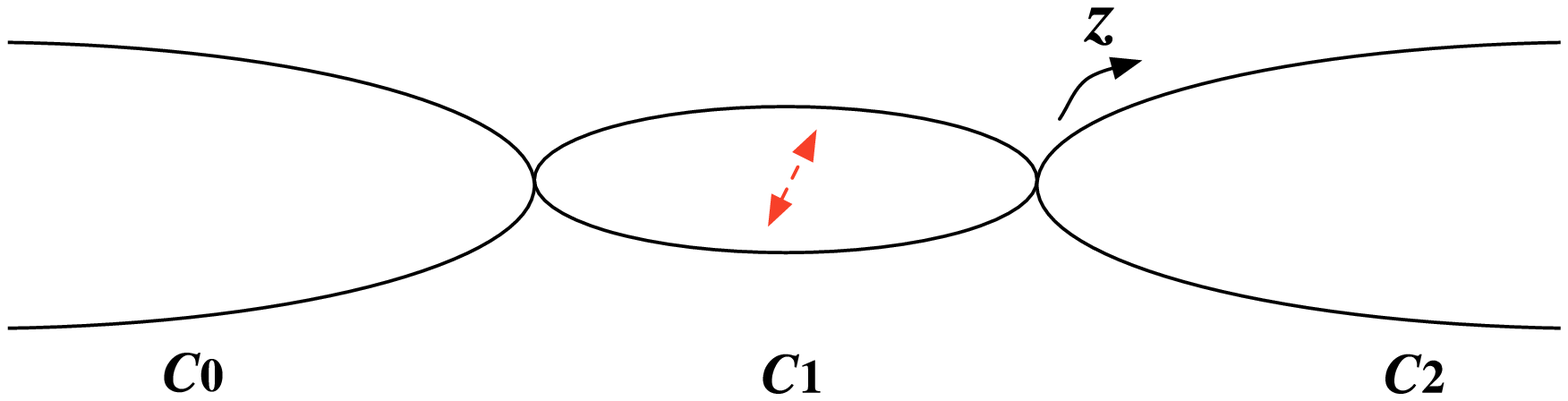}
\]
\caption{Another quiver, and the corresponding Taub-NUT in which its
Seiberg-Witten curve is embedded. 
Orientifolding fixes $C_{0}$ and $C_{1}$, but acts as multiplication by $-1$ on $C_{1}$.
\label{taub-so2}}
\end{figure}

The second example is the quiver drawn in Fig.~\ref{taub-so2};
$\USp(0)$ needs to be taken as a shorthand for the corresponding brane configuration.
This time $C_0$ and $C_2$ are both fixed by the orientifolding.
The Seiberg-Witten curve is 
\begin{multline}
0=v^6 y^4 + (c_1v^6+ u_1^{(2)} v^4 + u_1^{(4)}v^2 + u_1^{(6)}) y^3\\
+ v^2(c_2v^4+ u_2^{(2)} v^2 + u_2^{(4)}) y^2 
+ v^4(c_3v^2+ u_3^{(2)}) y
+ v^6.
\end{multline}
The intersection of this curve with $C_2$ parameterized by $z$
is determined by \begin{equation}
u_1^{(6)} + u_2^{(4)}z +  u_3^{(2)} z^2  + z^3 =0.
\end{equation}
Now, $C_2$ is a fixed locus of the M-theory orientifold,
and no M5-brane is wrapped on it. Therefore, 
the intersection needs to be a double zero
when an M5-brane intersects on it. 
This requires $u_1^{(6)}=0$ and $4u_2^{(4)}=(u_3^{(2)})^2$,
which leads to a simple pole in $\varphi_2$ and a double pole in $\varphi_4$
such that $4\varphi_4-(\varphi_2)^2$ only has a simple pole.
This property was crucial when we matched this defect of the $D_3$ theory
with the simple puncture of the $A_3$ theory.
In a similar manner, one finds that a quiver ending with groups 
$\SO(2k)$--$\USp(k-2)$ will allow a simple pole in $\varphi_k$ and 
a double pole in $\varphi_{2k}$, such that $4\varphi_{2k}-(\varphi_k)^2$ only has a simple pole.

\bibliography{All}{}
\bibliographystyle{utphys}
\end{document}